\documentclass{hep99}
\usepackage{epsfig}
\def\npb#1#2#3{#1 {\it Nucl. Phys. B} {\bf #2} #3}
\def\plb#1#2#3{#1 {\it Phys. Lett. B} {\bf #2} #3}
\def\prl#1#2#3{#1 {\it Phys. Rev. Lett.} {\bf #2} #3}
\def\prd#1#2#3{#1 {\it Phys. Rev. D} {\bf #2} #3}
\def\prep#1#2#3{#1 {\it Phys. Rep.} {\bf #2} #3}
\def\zpc#1#2#3{#1 {\it Z. Phys. C} {\bf #2} #3}
\def\nim#1#2#3{#1 {\it Nucl. Instr. Meth. } {\bf #2} #3}
\def\epm{$e^+e^-$}
\def\bbar{$B\overline{B}$}
\def\B{$B$}
\def\etal{{\it et al.}}
\def\Ell{{\cal L}}
\def\mt{$m_t$}
\def\mb{$m_b$}
\def\mc{$m_c$}
\def\pt{$p_t$}
\def\mQ{$m_Q$}
\def\dms{$\Delta m_s$}
\def\dmd{$\Delta m_d$}
\def\msbar{$\overline{MS}$}
\def\dec{\rightarrow}
\def\vcb{$|V_{cb}|$}
\def\vub{$|V_{ub}|$}
\def\vubcb{$|V_{ub}/V_{cb}|$}
\def\ups{$\Upsilon ({\rm 4S})$}
\newcommand{\sig}{\mbox{${ \cal S   }$}}
\newcommand{\sigbar}{\mbox{${ \overline{\sig} }$}}
\newcommand{\aerr}[3]   {\mbox{${{#1}^{+ #2}_{- #3}}$}}

\newcommand{\etaprkp}{\mbox{$\etapr K^+$}}

\newcommand{\etaprkz}{\mbox{$\etapr K^0$}}

\newcommand{\etaprpi}{\mbox{$\etapr\pi^+$}}

\newcommand{\etaprpiz}{\mbox{$\etapr\piz$}}

\newcommand{\etaprkstz}{\mbox{$\etapr K^{*0}$}}

\newcommand{\etaprkstp}{\mbox{$\etapr K^{*+}$}}

\newcommand{\etaprrhoz}{\mbox{$\etapr\rho^0$}}

\newcommand{\etaprrhop}{\mbox{$\etapr\rho^+$}}

\newcommand{\etak}{\mbox{$\eta K^+$}}

\newcommand{\etapi}{\mbox{$\eta\pi^+$}}

\newcommand{\etapiz}{\mbox{$\eta\piz$}}

\newcommand{\etakz}{\mbox{$\eta K^0$}}

\newcommand{\etakstz}{\mbox{$\eta K^{*0}$}}

\newcommand{\etakstp}{\mbox{$\eta K^{*+}$}}

\newcommand{\etarhoz}{\mbox{$\eta \rho^0$}}

\newcommand{\etarhop}{\mbox{$\eta \rho^+$}}

\newcommand{\retaKstp}{\mbox{$27.3^{+9.6}_{-8.2}\pm 5.0$}}

\newcommand{\retaKstz}{\mbox{$13.8^{+5.5}_{-4.4}\pm 1.7$}}

\newcommand{\retapKp}{\mbox{$80^{+10}_{-9}\pm8$}}

\newcommand{\retapKz}{\mbox{$88^{+18}_{-16}\pm9$}}

\newcommand{\piz}{\mbox{$\pi^0$}}
\newcommand{\acp}{\mbox{${{\cal A}_{\rm CP}}$}}
\newcommand{\goto}{\rightarrow}
\newcommand{\ra}{\rightarrow}
\newcommand{\calB}{\mbox{${\cal B}$}}

\newcommand{\etapr}{\mbox{$\eta^\prime$}}
\newcommand{\Bomegapi}{\mbox{$B^+\rightarrow\omega\pi^+$}}
\newcommand{\Bomegak}{\mbox{$B^+\rightarrow\omega K^+$}}
\newcommand{\Bomegah}{\mbox{$B^+\rightarrow\omega h^+$}}

\newcommand{\Bomegarhop}{\mbox{$B^+\rightarrow\omega \rho^+$}}
\newcommand{\Bomegarhoz}{\mbox{$B^0\rightarrow\omega \rho^0$}}
\newcommand{\Bomegakz}{\mbox{$B^0\rightarrow\omega K^{0}$}}
\newcommand{\Bomegapiz}{\mbox{$B^0\rightarrow\omega \pi^{0}$}}
\newcommand{\Bomegakstz}{\mbox{$B^0\rightarrow\omega K^{*0}$}}
\newcommand{\Bomegakstp}{\mbox{$B^+\rightarrow\omega K^{*+}$}}

\newcommand{\Bphik}{\mbox{$B^+\ra\phi K^+$}}
\newcommand{\Bphikz}{\mbox{$B^0\ra\phi K^0$}}
\newcommand{\Bphipi}{\mbox{$B^+\ra\phi\pi^+$}}
\newcommand{\Bphipiz}{\mbox{$B^0\ra\phi\pi^0$}}

\newcommand{\Brhozpi}{\mbox{$B^+\rightarrow\rho^0 \pi^+$}}
\newcommand{\Brhozk}{\mbox{$B^+\rightarrow\rho^0 K^+$}}
\newcommand{\Brhozpiz}{\mbox{$B^0\rightarrow\rho^0 \pi^0$}}
\newcommand{\Brhozkz}{\mbox{$B^0\rightarrow\rho^0 K^0$}}
\newcommand{\Brhompi}{\mbox{$B^0\rightarrow\rho^- \pi^+$}}
\newcommand{\Brhomk}{\mbox{$B^0\rightarrow\rho^- K^+$}}
\newcommand{\Bkstzpiz}{\mbox{$B^0\rightarrow K^{*0} \pi^0$}}
\newcommand{\Bkstzpi}{\mbox{$B^+\rightarrow K^{*0} \pi^+$}}
\newcommand{\Bkstppi}{\mbox{$B^0\rightarrow K^{*+} \pi^-$}}
\newcommand{\Bkstpk}{\mbox{$B^0\rightarrow K^{*+} K^-$}}
\newcommand{\Bkstzk}{\mbox{$B^+\rightarrow K^{*0} K^+$}}
\begin{document}
\title{Flavour Physics: the questions, the clues and the challenges }
\author{Marina Artuso }
\address{Department Physics, Syracuse University, Syracuse,
NY 13244, USA \\[3pt]
E-mail: {\tt artuso@phy.syr.edu}}
\abstract{Flavour physics addresses some of the questions for which the Standard 
Model does not provide a 
satisfactory and complete answer: the origin of the replication of the 
fundamental constituents and 
of their mass hierarchy. This paper reviews  
some of the theoretical approaches and the experimental strategies that can lead 
us 
to a more complete picture. Results included in this review are $|V_{cb}|=
0.0382 \pm 0.0032$, $|V_{ub}/V_{cb}|=0.085 \pm 0.023$  and a preliminary
measurement of the branching fraction of 
       $B\dec \pi^+\pi ^-=(0.47 ^{+0.18}_{-0.15}\pm 0.06)\times 10^{-5}$.} 
\maketitle

\section{Introduction}
The subject of flavour physics is a vast one and I am not going to review the 
entire body of 
present knowledge. Instead, I would
like to use a few examples to define the problems that we are struggling with 
and the
 strategies that we can use to reach a deeper
understanding of this elusive subject.

Our main puzzle was 
concisely and effectively put by I.I. Rabi upon the discovery of
the $\mu$: ``Who ordered that?" was his remark.
This question has three distinctive facets. We can start by asking the reason 
why there are so many
particles: this can be defined as the ``replication problem." Once we organize 
the particles in the 
Standard Model
families we have a triplication problem: ``Why are there three families?" 
Finally, the 
hierarchy in the mass spectrum 
of quarks and leptons spans several orders of magnitudes and may provide the 
most 
distinct clue towards an answer to Rabi's puzzling question.
Even if lepton masses and mixing parameters have recently attracted a lot of 
attention,
because of the evidence for $\nu$ oscillations from the 
Super-Kamiokande experiment, the quark sector 
will be the focus of my discussion. Masses and mixing in the lepton sector are 
reviewed in other
excellent contributions to these proceedings \cite{superk}.

In the Standard Model masses are produced via the 
Higgs mechanism through the Yukawa couplings. The Yukawa Lagrangian is given by:
\begin{equation}
\begin{array}{ll}
\Ell _{Yukawa} = & -G_1 [(\overline{L}_u\overline{\Phi})u_R 
+\overline{u}_R(\overline{\Phi}^{\dagger }L_u)]\\ 
~~ & - G_2 [(\overline{L}_u\overline{\Phi})d_R 
+\overline{d}_R(\overline{\Phi}^{\dagger }L_u)] \\ 
~~ & -G_3 [(\overline{L}_u\overline{\Phi})s_R 
+\overline{s}_R(\overline{\Phi}^{\dagger }L_u)] \\
~~ & -G_4 [(\overline{L}_c\overline{\Phi})c_R 
+\overline{c}_R(\overline{\Phi}^{\dagger }L_c)] \\ 
~~ & -G_5 [(\overline{L}_c\overline{\Phi})d_R 
+\overline{d}_R(\overline{\Phi}^{\dagger }L_c)]\\
~~ & -G_6 [(\overline{L}_c\overline{\Phi})s_R 
+\overline{s}_R(\overline{\Phi}^{\dagger }L_c)],\\
\end{array}
\end{equation}
where $L$ identifies a left-handed SU(2) doublet and the subscript $R$ 
identifies a right-handed
fermion. 
These 6 complex Yukawa couplings are related to 10 independent physical 
quantities, the six quark
 masses and  four quark
mixing parameters. The latter can be described by three 
Euler-like angles and an imaginary phase. These parameters are
not predicted by the Standard Model, but 
are fundamental constants of nature that 
need to be extracted from experimental data.

Many theoretical models have tried to uncover a more fundamental 
explanation for flavour. For example, some 
of the many variations of Supersymmetry \cite{susy} incorporate 
the known hierarchy of quark masses and mixing parameters. 
In addition, the replication problem has been addressed 
by postulating a new deeper level of matter \cite{preons}. In this approach, the 
multitude
of quarks can be understood as a sort of periodic table of the composite 
structures 
that are indeed bound states of more fundamental particles.
In addition, a geometrical origin \cite{string} of flavor has been proposed.

An additional feature of the fundamental interactions explored in
flavour physics is $CP$ violation. The only experimental evidence 
  has been obtained  studying neutral $K$ decays. On the
other hand, $CP$ violation is  crucial to our understanding of
the history of the universe.  In particular, it is a necessary ingredient of
our understanding of the origin of the matter dominated universe 
\cite{sakharov}. A $CP$ violating
phase is naturally incorporated in the Standard Model within the Cabibbo-
Kobayashi-Maskawa matrix. 
Thus several models attempt to explain the baryon asymmetry of the universe as 
due 
to a $CP$ violating process occurring at the scale of the electroweak symmetry 
breaking.
A rough order of magnitude estimate of the expected effect of 
the CKM induced $CP$ violation on the baryon asymmetry can be obtained by
constructing a variable $d_{CP}$ that incorporates all the 
features of the expected CKM phase: it vanishes when any pair of quarks is 
degenerate in mass 
and when any CKM angle vanishes
because of the so called ``GIM'' (Glashow, Iliopoulos, Maiani) cancellation.   
$d_{CP}$ is defined as:
\begin{equation}
\begin{array}{ll}
d_{CP} = & \sin {\theta _{12}} \sin {\theta _{23}}\sin {\theta _{13}} \sin 
{\delta _{CP}}\\ 
~~&(m_t^2-m_c^2)(m_t^2-m_u^2)(m_c^2-m_u^2)\\
~~&(m_b^2-m_s^2)(m_b^2-m_d^2)(m_s^2-m_d^2),\\
\end{array}
\end{equation}
where $\theta _{ij}$ are three real ``Euler-like''
angles defining the CKM matrix together with the imaginary phase $\delta _{CP}$.

The $d_{CP}$ parameter that we have just defined is a dimensional quantity, 
it is conceivable \cite{farrar}
that the natural normalization parameter to transform it into
a pure number  is the temperature at which the electroweak
symmetry breaking occurred. Thus the figure of merit of the strength of the CKM
induced CP violating effect is given by:
\begin{equation}
d_{CP}^T = d_{CP}/kT_{ew}^{12}  \approx 10^{-18},
\end{equation}
where $T_{ew}$ represents the temperature at the time the electroweak symmetry 
breaking occurred and
$k$ is the Boltzmann constant.
This suggests that CKM CP violation is an effect too small to account for the 
known baryon asymmetry of the universe,
\begin{equation}
\left|\frac{N_{B}-N_{\overline{B}}}{N_{B}+N_{\overline{B}}}\right| _{t\approx 
10^{-6} s}\approx 
\left|\frac{N_B}{N_{\gamma }}\right| _{now}\approx 10^{-10}.
\end{equation}
This discrepancy is very qualitative in nature and may have a number of 
explanations. However a
very tantalizing hypothesis is the presence of additional $CP$ violating phases 
produced by
mechanisms beyond the Standard Model. Thus, the experimental exploration of $CP$ 
violation observables 
has a good chance to
uncover evidence for new physics.
\section{Quark Masses}
The hierarchy of the quark masses is very puzzling. The mass interval between 
the lightest 
and the heaviest quark  spans the huge interval between a few MeV and hundreds 
of GeV. The 
determination of these masses has many challenges and a good illustration of the 
different roles 
played by QCD in different hadronic processes. The parameter that defines how 
QCD affects our ability of 
measuring fundamental properties is the mass scale $\Lambda _{QCD} \approx 500$ 
MeV. Thus, for very 
heavy mass scales QCD corrections are not very important and can be evaluated 
using perturbative 
methods. As the mass decreases, non-perturbative effects become more important 
and the extrapolation 
process becomes more uncertain. This interplay will be illustrated below.
\subsection{The top quark mass}

The top quark is the heaviest and thus the least affected by the strong 
interaction. In fact, it is so heavy 
that it has the unique property of decaying before hadronizing in a top-flavored 
hadron. It was discovered 
at Fermilab \cite{topdiscovery}, at the Tevatron proton-antiproton collider. Its 
properties have been 
studied by  CDF and D0. In 
particular, one of the finest achievements by these groups is a very accurate 
determination of the quark 
mass \mt . Several different techniques are used. Recently, these two 
experiments have 
formed a working group to obtain a world average for \mt , considering the 
correlations in their 
measurements.

%\end{verbatim*}
The top quark decays into a $b$ quark and a $W$ boson with a probability close 
to one. In some 
cases this $b$ quark
undergoes semileptonic decays. In both CDF and D0 the dominant contribution to 
the \mt\ measurement comes from
samples containing one high \pt\ lepton and one jet. CDF also uses a sample
including only hadrons in the final state, whereas D0 has a secondary data set 
containing two
leptons.

The value of the measured top quark mass is $(174.3 \pm 3.2  \pm 4.0)$ GeV. This 
is the heaviest 
measured mass of a fundamental particle. The interest in studying the properties 
of the top
quark is largely motivated by the unique closeness of $m_t$ to the mass defining 
the scale of the 
electroweak symmetry breaking. The mass \mt\ is related to the vacuum 
expectation value of the
Higgs field through the relationship:
\begin{equation}
m_t = \lambda _t v/\sqrt{2},
\end{equation}
where $\lambda _t$ is the Yukawa coupling and $v$ is the vacuum expectation 
value of the Higgs
field. Note that the Yukawa coupling is of order one. This property can be 
accomodated in
a natural way in some supersymmetric theories that encompass a very heavy top 
quark at the GUT
scale \cite{susy-guts}.
\begin{figure}[hbt]
\begin{center}
\epsfig{figure=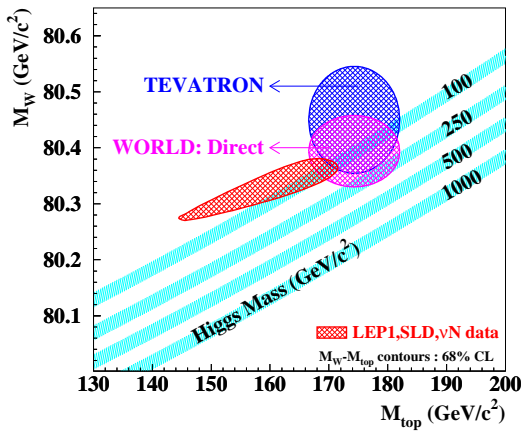,width=2.7in}
\caption{ Relation between the W and top masses for some values
    of the Higgs mass \cite{galtieri}. The cross hatched area to the 
left represents the
    allowed region from an overall Standard Model fit. The other two regions
    are obtained from direct measurements of M$_W$ and M$_{top}$.} 
\label{fig:mhiggs}
\end{center}
\end{figure}
 
The measured $W$ and top masses can give some constraint on the mass scale for 
the last particle in the 
Standard Model to elude our discovery, the Higgs scalar. Fig.~\ref{fig:mhiggs} 
shows a summary of the constraints
on the Higgs mass from the $W$ mass measurements at LEP II and the combined 
$m_W$ and $m_t$ at the Tevatron.
These data favor a light Higgs, leading to the hope that we are on the verge of 
learning more 
about the mechanism of electroweak symmetry breaking.

\subsection{The $b$ quark mass}
The interest in an accurate
measurement of the $b$ quark mass has a two-fold motivation. On the one hand its 
measurement helps to
pin down the quark mass hierarchy as discussed before. On the other
hand, renewed attention to \mb\ has been motivated by the 
suggestion that the CKM matrix elements \vcb\ and \vub\ 
 can now be obtained from inclusive measurements with great accuracy using 
the so 
called ``Heavy Quark Expansion'' \cite{ope}. A precise determination
 of $m_b$ is a ``sine qua non'' for most inclusive methods to study $B$ 
decays.
Since the mass of the beauty quark is about 5 GeV, thus its derivation from 
experimental observables is
affected by the strong interaction to a much higher degree that in the $t$ quark 
case.
In fact, before we discuss any theoretical estimate or experimental 
determination of the $b$ quark mass
we need to define its meaning carefully. 
The widely used  ``pole mass,'' namely the parameter appearing in the quark 
propagator, has
been demonstrated to be inadequate \cite{bigietal} for an accurate description 
of 
$b$ quark phenomenology. These 
authors point out the existence of an infrared renormalon generating a factorial
divergence in the high-order cofficients in the $\alpha _S$ series producing an 
intrinsic uncertainty
of the order $\Lambda _{QCD}/m_Q$. 

An alternative, that appears to be quite a good candidate to describe $b$ quark 
properties in a regime
where perturbative effects are important, is the \msbar\ mass 
$\overline{m}_b(\mu )$. This 
is a short distance mass, and therefore does not contain ambiguities of the 
order of $\Lambda _{QCD}$ like
the pole mass. This definition is adequate for observables where perturbative 
effects are dominant,
such as recent measurements of the $b$ quark mass at the $Z^0$ energy.
As this quantity is not defined for scales $\mu$ below $m_b$, alternative 
definitions of the running mass
have been proposed that can be normalized at a scale $\mu$ smaller than $m_b$. 
Most of the authors refer
to the mass defined with this procedure as ``kinetic mass.'' 

Table~\ref{bmass} summarizes the most recent theoretical evaluations of $m_b$. 
The calculations using
the $m_{kin}$ prescription or an equivalent subtraction scheme are in 
closer agreement. However other evaluations of the pole mass and the 
$\overline{MS}$ mass 
have shown frequent disagreement in the methodology of assessing errors. In 
addition,
recent lattice data \cite{yashimoto} seem to imply an even lighter \mb . It is 
clear that
there is still work to be done before quantities sensitive to \mb\ can be 
determined
accurately. 

\begin{table}
\begin{center}
\caption{\label{bmass} Theoretical estimates of $m_b$ (GeV/c$^2$). The first 
column gives the
theoretical predictions for the pole mass $m_b$, the second column the predicted 
value of the $\overline{MS}$ mass $\overline{m}_b(m_b)$ at
the scale of the $b$ meson mass and the third column corresponds to the 
prediction for the {\it kinetic} or {\it potential subtracted}
mass defined at a scale $\mu \approx 1$ GeV.}
\begin{tabular}{lccc} 
\br
$m_b$ & $\overline{m}_b$ & $m_{kin}$ & Ref. \\
\hline
\multicolumn{4}{c}{\it QCD sum rule evaluations}\\
\hline
-- & $4.20\pm 0.1$ & $ 4.56\pm 0.06$ &\cite{my98}  \\
-- & $4.20\pm 0.06$ &$4.71\pm 0.06$  & \cite{h98}\\
  $4.97\pm 0.17$ & $4.26\pm 0.12$ & $4.60 \pm 0.18$ &\cite{bs} \\
 $4.75 \pm 0.04$ & -- & -- & \cite{kpp} \\ \hline
\multicolumn{4}{c}{\it $\Upsilon$ sum rule evaluations}\\
\hline
$4.604\pm 0.014$ & $4.13\pm 0.06$& --& \cite{jp978}\\
$5.015 ^{+0.110}_{-0.070}$ & $4.453 ^{+0.050}_{-0.032}$& --   &\cite{py98}
 \\
\hline
\multicolumn{4}{c}{\it Lattice QCD evaluations}\\
\hline
 $5.0\pm 0.2$ & $4.0\pm 0.1$ & -- &  \cite{deta} \\
 -- & $4.41\pm 0.15$ &-- & \cite{ms98} \\
\br
\end{tabular}
\end{center}
\end{table}

\section{Quark Mixing}
In the framework of the Standard Model the gauge bosons, $W^{\pm}$, 
$\gamma$ and 
$Z^o$ couple to  
mixtures of the physical $d,~ s$ and $b$ states. This mixing is described
by the Cabibbo-Kobayashi-Maskawa (CKM) matrix:
\begin{equation}
V_{CKM} =\left(\begin{array}{ccc} 
V_{ud} &  V_{us} & V_{ub} \\
V_{cd} &  V_{cs} & V_{cb} \\
V_{td} &  V_{ts} & V_{tb}  \end{array}\right).
\end{equation}
A commonly used approximate parameterization was originally proposed by 
Wolfenstein \cite{wolf}. 
It reflects the hierarchy between the magnitude of matrix elements 
belonging to different 
diagonals. It is:
\begin{equation}
\scriptsize
{\begin{array}{ccc}
1-\lambda^2/2 &  \lambda & A\lambda^3(\rho-i\eta(1-\lambda^2/2)) \\
-\lambda &  1-\lambda^2/2-i\eta A^2\lambda^4 & A\lambda^2(1+i\eta\lambda^2) \\
A\lambda^3(1-\rho-i\eta) &  -A\lambda^2& 1  
\end{array}} 
\end{equation}

\begin{figure}[hb]
%\vspace{2cm}
\centerline{\epsfig{figure=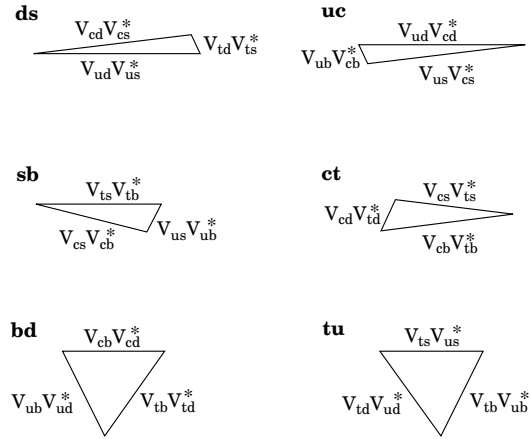,height=2.3in}}
%\vspace{1cm}
\caption{\label{six_tri}The six CKM triangles. The bold labels, i.e {\bf ds} 
refer to the rows or columns used in the unitarity relationship.}
\end{figure}

The CKM matrix must be unitary and the relation between elements of different
rows dictated by this property can be graphically represented as
so called `unitarity triangles'.   Fig.~\ref{six_tri} shows the 6 unitarity 
triangles
that can be constructed to check this property. The expected lengths of the 
sides
are suggested. The bottom two triangles give a hint on the optimal strategy 
to test the CKM sector of the Standard Model. 
The three sides are of comparable length and thus all the angles and sides are 
more amenable to experimental measurement. The CKM matrix is characterized by 
four independent parameters. Aleksan, Kayser and London \cite{alk} pointed out 
that
we can express the CKM matrix as a function of four phases that can be 
determined
by $CP$ observables. These phases are:
\begin{eqnarray}
\beta=arg\left(-\frac{V_{tb}V_{ts}^{\star }}{V_{cb}V_{cd}^{\star }}\right) \\
\gamma =arg\left(-\frac{V_{ub}^{\star }V_{ud}}{V_{cb} ^{\star }V_{cd}}\right) \\
\chi = arg\left(-\frac{V_{cs}^{\star }V_{cb}}{V_{ts}^{\star }V_{tb}}\right) \\
\chi ^{\prime }=arg\left(-\frac{V_{ud}^{\star }V_{us}}{V_{cd}^{\star 
}V_{cs}}\right) 
\end{eqnarray}
 Silva and 
Wolfenstein \cite{wol-si} emphasized that we can use these
phases to identify new physics effects in $B$ 
decays. The details of an 
experimental program capable of making these 
measurements with the needed accuracy will be discussed below.

The experimental information presently available consists of $CP$ violation 
observables
in the $K$ system, discussed by G. Buchalla \cite{Gerhard} in these
proceedings, and by a variety 
of constraints on the sides of the unitarity triangles. Very often the dominant 
uncertainty
in the extraction of the magnitude of the CKM parameters from the data is the 
relationship
between experimental observables and the quark mixing parameter to be measured.
This relationship is governed by a matrix element involving strong interaction 
effects. The challenges
and possible pitfalls in the evaluation of the relevant matrix elements are 
discussed below.

\subsection{The determination of $V_{cb}$ from the decay $B\dec D^{\star } \ell 
\overline{\nu}$.} 
Heavy Quark Effective Theory (HQET) \cite{isgur} has been a very important 
breakthrough in 
our understanding of $B$ meson decays. It is an effective theory that 
gives very definite predictions in the limit of infinite quark masses. In 
addition, 
corrections for finite quark masses can be accounted for  with a systematic 
study
of non-perturbative effects using a 1/\mQ\ expansion, where $m_Q$ represents 
the mass of the heavy quarks involved in the process. One of the first 
implication of the theory
\cite{luke} has been the advantage offered by the decay $B\dec D^{\star } \ell 
\nu$. The differential
decay width $\Gamma (B\dec D^{\star } \ell \overline{\nu})$ is a function of the 
invariant 4-velocity
transfer $w=\vec{v}\cdot \vec{v^{\prime}}$, where $\vec{v}$ and 
$\vec{v^{\prime}}$
are the 4-velocities of the incoming and outgoing heavy hadron respectively. It 
is given by:
\begin{eqnarray}
\frac{d\Gamma}{dw}= {\cal K}(w){\cal F}^2(w)|V_{cb}| ^2,\\
{\cal F}(w) = {\cal F}_{D^ {\star }}(1){\cal G}(w)
\end{eqnarray}
where ${\cal K}(w)$ is a known phase space factor, ${\cal G}(w)$ is the shape of 
the Isgur-Wise function and
${\cal F}_{D^ {\star }}(1)$ is a normalization factor that is predicted from 
this effective theory.
It is generally expressed as:
\begin{eqnarray}
{\cal F}(1) =\eta _A ( 1 + const 
\times  \frac{\Lambda _{QCD}^2}{m_Q^2}+ \cdots )
\\ \nonumber \equiv
\eta _A (1+\delta _{1/m_Q^2})
\end{eqnarray}
Note, the $1/m_Q$ term vanishes.
The parameter $\eta _A$ is the short distance correction arising from the finite 
renormalization of the flavour changing
axial current at zero recoil and $\delta _{1/m_Q^2}$ parametrizes the second 
order term (and higher) order 
corrections in the $1/m_Q$ expansion including the effects of finite heavy quark 
masses.
There has been a lot of theoretical activity on both $\eta _A$ and $1/m_Q^2$ 
corrections. The $\eta _A$ factor
has been calculated by Czarnecki at two loop order \cite{cza} to be $0.960\pm 
0.007$. The $1/m_Q^2$ correction has been
evaluated in different approaches and is more subjected to the author bias 
because it involves non-perturbative effects
\cite{variety1}, \cite{variety2}, \cite{variety3},
\cite{variety4}, \cite{variety5}. 
These different estimates have been combined into \cite{babar}:
\begin{equation}
{\cal F}(1)=0.913\pm 0.007\pm 0.024 \pm 0.011,
\end{equation}
where the first error accounts for the remaining perturbative uncertainty,
the second one reflects the uncertainty in the calculation of the $1/m_Q^2$
corrections and the third one gives an estimate of the higher-order power 
corrections.
 A very interesting recent development
is that Lattice Gauge theory has produced a preliminary
result on ${\cal F}(1)$. They obtain \cite{simone}, \cite{dlnu} :
\begin{equation}
{\cal F}(1)=0.935\pm 0.022 ^{+0.008}_{-0.011}\pm 0.008 \pm 0.020,
\label{normds}
\end{equation}
where the first error is statistical, the second is due to the uncertainty in 
the 
quark masses, the third is the uncertainty in radiative corrections beyond 1 
loop and
the last represent $1/m_Q^3$ effects. Errors not yet estimated include the 
sensitivity to
the lattice spacing adopted and to the quenching approximation.
The calculation is in its initial stage, but this is a promising hint that 
an accurate value for
${\cal F}(1)$ will be available soon.

The present status of the experimental determination of the product 
${\cal F}(1) V_{cb}$ is summarized in Table \ref{vcbexcl}.
\begin{table}
\begin{center}
\caption{\label{vcbexcl}Determination of \vcb\ 
from $B\dec D^{\star}\ell \overline{\nu }$. 
The first error in the \vcb\ estimate is the experimental error, obtained
adding in quadrature statistical and systematic errors. The last error is due to 
theoretical 
uncertainties in ${\cal F} (1)$.}
\begin{tabular}{lll} 
\br
Experiment & ${\cal F}(1) V_{cb}\times 10^{-3}$ & $V_{cb}\times 10^{-3}$ \\
\hline
ALEPH \cite{alds}
  & $31.9\pm 1.8\pm 1.9$ & $34.9 \pm 2.9 \pm 1.6$   \\
DELPHI\cite{del99} & $37.7\pm 1.7 \pm 1.7$ &$ 41.3 \pm 2.6 \pm 1.9$\\
OPAL\cite{opl96}
& $32.8\pm 1.9 \pm 2.2$ & $35.9 \pm 3.2 \pm 1.5 $\\
\hline
CLEO $\dagger$ & $36.1 \pm 1.9 \pm 1.9$ & $39.5 \pm 2.9 \pm 1.8$ \\ 
\hline
AVERAGE &  $34.8\pm 0.91\pm 0.95$& $38.2\pm 1.4 \pm 1.8 $\\
\hline
\end{tabular}
\end{center}
%\begin
$\dagger$ The published CLEO result \cite{dscleo} has been scaled up by
3\% according to the calculation performed by Stone \cite{albuquerque} to 
account
for the curvature of the form factor.
%\end{center}
\end{table}

If we average these results, we obtain $V_{cb}=(38.2 \pm 1.40 _{exp} 
+1.8_{th})\times
10^{-3}$.
The experimental error is obtained by adding in quadrature the statistical and 
systematic errors.  The 
theoretical
error is based on Eq.~\ref{normds}. 

\subsection{The determination of \vcb\ from inclusive semileptonic decays.}

\begin{figure}
\begin{center}
\epsfig{file=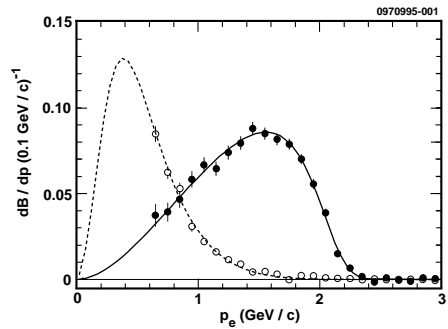,width=2.7in}
 \caption{Spectra of electrons from 
$B\dec  X e \overline{\nu }$ (filled circles) and $b\dec  c \dec y l 
\overline{\nu}$ 
(open circles) identified with
lepton tags. The curves show the best fit to the modified ISGW model, with 23\% 
$B\dec D^{\star \star}\ell \overline{\nu}$.}
\label{sldile}
\end{center}
\end{figure}

An alternative determination of $V_{cb}$ has been obtained using the inclusive
semileptonic branching fraction. This quantity has been studied both at the
\ups\ from CLEO and ARGUS and at LEP. The most accurate measurement at 
the \ups\
has been reported by CLEO from a small subset of their full data sample 
\cite{dilep} using a lepton tagged sample. This method of extracting the
branching fraction is less model dependent as it can measure a larger portion
of the electron spectrum as shown in Fig.~ \ref{sldile}.

 All the four LEP 
experiments reported measurements of the semileptonic branching fractions. 
The experimental information is summarized in Table \ref{semil}. The
measured semileptonic widths, obtained using the most recent average values
for the $b$ quark  and the $B$ meson lifetimes, are shown also. Note that the
partial width measurements from LEP and the \ups\ are still mildly
inconsistent. In addition, the 
systematic
errors in different LEP experiments are evaluated using different methods and
assumptions.
A working group has been established to address this issue \cite{lepwg}, but 
their
work is still in progress. As the dominant error in these analyses is 
systematic, until
these issues are settled it appears premature to use a weighted average of these
results.

\begin{table}
\begin{center}
\caption{\label{semil}Summary on semileptonic width data from CLEO and LEP.
The lifetimes $\tau _B=1.605\pm 0.021$ ps and $\tau_b=1.564\pm 0.014$ ps are 
used to obtain
$\Gamma _{SL}$ for CLEO and LEP respectively.}
\begin{tabular}{lll}
\br
Exp. &  ${\cal B}_{SL}$(\%) & $\Gamma _{SL}$ \\
  &  & $\times 10^{10} {\rm s}^{-1}$ \\
\hline
ALPH \cite{alsl} & $11.01\pm 0.10\pm 0.30$ & $7.04 {\scriptsize \pm 0.21} $ \\
DPHI \cite{delsl} & $10.65\pm 0.11 \pm 0.23 ^{+0.43}_{-0.27}$ & $6.81 
^{+0.32}_{-0.23}$ \\
L3\cite{l3sl}& $10.16 \pm 0.13 \pm 0.29 $ & $6.50 {\scriptsize \pm 0.20}$ \\
OPAL\cite{opalsl} & $10.83 \pm 0.10 \pm 0.20 ^{+0.20}_{-0.13}$ & $6.92 
^{+0.19}_{-0.17}$\\
\hline
CLEO\cite{dilep} & $10.49 \pm 0.17 \pm 0.43$ & $6.54 {\scriptsize \pm 0.29}$\\
\br
\end{tabular}
\end{center}
\end{table}

Recently  proponents
of the heavy quark expansion (HQE) \cite{ope-sl} have suggested that inclusive
quantities are the most promising avenue to perform a precise determination
of the CKM parameters. The semileptonic width is related to \vcb\ and
to the matrix element calculated by HQE through:
\begin{equation}
\begin{array}{ll}
|V_{cb}| = & 0.0411 \sqrt {{\cal B}(B\dec X_c\ell \overline{\nu})/0.105} \\
~~ &\times \sqrt{1.55/\tau _B ({\rm ps})}\\ 
~~ &\left[1-0.024\left(\frac{\mu_{\pi }^2-0.5}{0.1}\right)\right]\times \\ 
~~ & (1\pm 0.01 (pert.)\pm 0.01 (m_b) \\
~~ & \pm 0.015 (1/m_b^3))
\end{array}
\end{equation}
where $\mu _{\pi}^2$ is a kinetic energy contribution which is a gauge-covariant
extension of the square of the $b$ quark momentum inside the 
heavy hadron, $\mu _G$ is the chromomagnetic matrix element defined to a 
given order in perturbative theory. Alternative formulations of this 
relationship
have been proposed, either using slightly different parameterizations of the 
correction 
terms \cite{falk-incl}, or relating the semileptonic width to the $\Upsilon$ 
mass
in order to reduce the uncertainties associated with quark masses \cite{zoltan}. 
If, for illustration purpose, we use the CLEO semileptonic width to extract 
\vcb\ with this
method, we obtain $|V_{cb}| = 0.041 \pm 0.001_{exp} \pm 0.0025_{th}$. 

 One very important concern is the dependence upon the quark
masses. Bigi \cite{bigiwg}
assumes an uncertainty of 10\%. Recent theoretical estimates of $m^{kin}_b$
seem to support this view, but more experimental data confirming
this picture is needed.  The
semileptonic width in the heavy quark expansion is dependent upon \mc\ and
\mb\ though the relationship:
\begin{equation}
\Gamma(B\dec X_c\ell \overline{\nu}) \propto G_F^2(m_b-m_c)^2m_b^3
\end{equation}

The difference $m_b-m_c$ can be inferred according to several 
authors \cite{hqe}
through the relationship:
\begin{equation}
\begin{array}{ll}
m_b - m_c = & <M_B> - <M_D> + \\
~~ & \mu ^2_{\pi }\left(\frac{1}{2m_c}-\frac{1}{2m_b}\right) + \\
~~&{\cal O}(1/m_{b,c}^2),
\end{array}
\end{equation}
where $<M_B>$ and $<M_D>$ represent the spin averaged beauty and charm meson 
masses.
This formula gives $m_b - m_c \approx 3.5$ GeV, corresponding to an 
uncomfortably low value of
the $c$ quark mass \cite{bigiwg}.

Finally it is hard to give a reliable estimate of the possible quark hadron 
duality
violation in inclusive decays. In fact there are some arguments that suggest a
potential for significant violations of this assumption \cite{nathan}.
Until these issues are resolved, an average value of \vcb\ based on the 
$D^{\star }\ell \overline{\nu}$ method 
and the inclusive semileptonic widths does not seem appropriate.
 
Additional experimental
constraints are crucial to achieve a better understanding of these sources of 
errors. The study of the
moments of the lepton spectrum and of the hadronic mass spectrum provides in 
principle an additional
constraint that allows the extraction of some of the theoretical parameters 
described above from
data. The moments of the hadronic mass spectrum are defined as:
\begin{equation}
<(s_H-\overline{m}_D^2)^n)>\equiv\frac{1}{\Gamma _{SL}}\int ds \frac{d\Gamma 
}{ds}(s- \overline{m}_D^2)^n,
\end{equation}
where $s$ represent the hadronic mass recoiling against the lepton-neutrino pair 
and $\overline{m}_D$ represents the
spin averaged mass of the $D$ meson.
If we use the heavy quark expansion result for $\Gamma _{SL}$, we obtain 
for the 
first moment \cite{falk-incl} :
\begin{equation}
\begin{array}{ll}
<(s_H-\overline{m}_D^2)> =& m_B^2 [ 0.051 \frac{\alpha _S}{\pi} +\\ 
~~ & 0.23 \frac{\overline{\Lambda}}{m_B}(1+0.43\frac{\alpha _S}{\pi})\\ 
~~ &+0.26\frac{\overline{\Lambda }^2}{m_B^2} +1.01 \frac{\lambda _1}{m_B^2}\\
~~ & -0.31
 \frac{\lambda _2}{m_B^2}]
\end{array}
\end{equation}
where the two matrix elements $\lambda _1$ and $\lambda _2$, characterizing the 
corrections of
order $\Lambda _{QCD}/m_b^2$ are given by:
\begin{eqnarray}
\lambda _1 =-\frac{1}{2m_B}<B|\overline{h}_{\nu }(i\vec{D})^2 h_{\nu}| B>, \\
\lambda _2 =\frac{1}{3}<B| \overline{h}_{\nu }\frac{g}{2}\sigma _{\mu 
\nu}G^{\mu\nu}h_{\nu}| B>, 
\end{eqnarray}
where  $h_{\nu}$ is the heavy quark field in the effective theory with velocity 
$\vec{v}$.
The parameter $\overline{\Lambda}$ is related to the $b$ quark mass through the
relationship:
\begin{equation}
m_B =m_b +\overline{\Lambda} -\frac{\lambda _1+3\lambda _2}{2m_b}.
\end{equation}
The $B^{\star}-B$ mass splitting determines $\lambda _2=0.12\ 
{\rm GeV}^2$. The parameters $\overline{\Lambda}$
and $\lambda _1$ can be extracted from the measured moments. Each measured 
moment corresponds
to a band in the $\lambda _1 -\overline{\Lambda}$ plane. The CLEO collaboration 
\cite{poling-vanc} measured first and second moments of the hadronic
mass have a region of intersection that implies:
\begin{eqnarray}
\overline{\Lambda} = 0.33\pm 0.02 \pm 0.08 {\rm GeV}\\
\lambda _1 = -0.13 \pm 0.01 \pm 0.06 (\rm (GeV/c)^2.
\end{eqnarray}

\begin{figure}[hbt]
\vspace{-5mm}
\centerline{\epsfig{figure=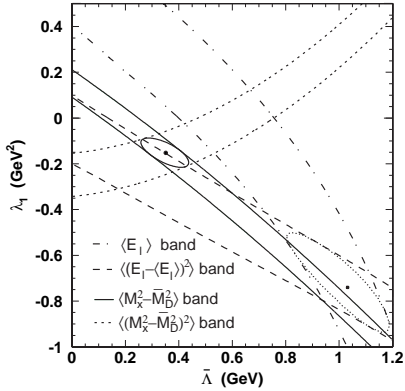,width=2.1in}}
\vspace{5mm}
\caption{\label{moments}Bands in $\overline{\Lambda}-\lambda_1$ space found
by CLEO in analyzing first and second moments of hadronic mass squared and
lepton energy. The intersections of the two moments for each set determines
the two parameters. The one standard deviation error ellipses are shown.}
\end{figure}

The two curves are shown in Fig.~\ref{moments}, that shows also the constraints 
derived from a similar analysis
of the experimental moments of the lepton spectrum. It can be seen that the two 
sets of constraints do not intersect at
a common point as they should. These data are still preliminary and we cannot 
draw definite conclusions from them.
However until this discrepancy is resolved, this 
is yet another reason not to use
an average value for $V_{cb}$ based
on the inclusive and exclusive determinations.

\subsection{The determination of \vub\ from exclusive charmless semileptonic 
decays}
Unfortunately, in the case of \vub\ HQET does not help to 
normalize the relevant 
form factors. A variety of
calculations of such form factors exist, based on lattice gauge theory 
\cite{ukqcd}, light cone sum rules  (LCSR) \cite{lcsr},
and quark models \cite{isgw}. The CLEO 
collaboration has reported the first
convincing evidence for the decays $B\dec \rho \ell \overline{\nu}$ and $B\dec 
\pi \ell \overline{\nu }$ \cite{buprl}. 
CLEO has recently reported a measurement of the decay $B \dec \rho \ell 
\overline{\nu }$ with a different technique and a bigger
data sample \cite{summerrho}. They have used several 
different models to extract 
the value of \vub . Their results are summarized 
in Table \ref{vubrho}.

\begin{table}
\begin{center}
\caption{\label{vubrho} Values of \vub\ using  $B\dec \rho \ell \overline{\nu }$ 
and some theoretical models. The \vub\ data
include the results of a recent CLEO analysis \cite{summerrho} and a previous 
CLEO result on exclusive charmless semileptonic
decays \cite{rhoprl}. The average \vub\ includes an additional contribution 
representative of the theoretical uncertainty
in the measurement.}
\begin{tabular}{ll}
\br
Model & \vub\ ($\times 10^{-3}$)\\
\hline
UKQCD \cite{ukqcd} & $3.32\pm 0.14 ^{+0.21}_{-0.26}$ \\
LCSR \cite{lcsr} & $3.45 \pm 0.15 ^{+0.22}_{-0.31}$ \\
ISGW2 \cite{isgw} & $ 3.24 \pm 0.14 ^{+0.22}_{-0.29}$ \\
Beyer-Melikhov \cite{bm} & $ 3.32 \pm 0.15 ^{+0.21}_{-0.30}$ \\
Wise/Ligeti \cite{wl} & $2.92 \pm 0.13 ^{+0.19}_{-0.26}$ \\
\hline
Average & $3.25\pm 0.14 ^{+0.21}_{-0.29}\pm 0.55$ \\
\br
\end{tabular}
\end{center}
\end{table}

The first three calculations are based on quark models and their uncertainties 
are guessed to be in the 25-50\%
range in the rate, corresponding to a 12.5-25\% uncertainty for \vub . The other 
approaches, light cone sum rules and lattice
QCD, estimate their errors in the range of 30\%, leading to a 15\% error in \vub 
. We can conclude that the average value
of \vub\ extracted with this method is $| V_{ub}| = (3.25 \pm 0.14 ^{+0.21}_{-
0.31}\pm 0.5)\times 10^{-3}$. This corresponds
to a value of $|V_{ub}/V_{cb}| = 0.085 \pm 0.023$. The statistical and 
systematic errors have been added in quadrature
and the theoretical error has been added linearly to be conservative. Note that 
the theoretical error is somewhat arbitrary,
as the spread between the models considered does not necessarily 
represent the uncertainty in $\Gamma _u^{th}$ nor may it properly reflect the
effect of the lepton momentum cut of $>2.3$ GeV/c used in the analysis.

An alternative
 approach used by CLEO has been the investigation of the inclusive lepton 
spectrum beyond the kinematic
endpoint of $B\dec X_c \ell \overline{\nu}$ decays \cite{cleoinc}. This method 
gave the first unambiguous evidence 
for charmless semileptonic decays, however its use to extract \vub\ is plagued 
by several theoretical
uncertainties. In addition to the errors discussed above, there is the 
additional problem that very few 
models can be used to relate $\Gamma _u ({\rm endpoint})$ with $\Gamma _u$. In 
fact, most of the models 
discussed above study only final states like $\rho \ell \overline{\nu}$ and $\pi 
\ell \overline{\nu}$, whereas
it is likely that $\Gamma _u$ is composed of several different hadronic final 
states. The Operator Product
Expansion (OPE) cannot give reliable predictions either, because when the 
momentum region considered is of the
order of $\Lambda _{QCD}$ an infinite series of terms in this expansion may be 
relevant. Nonetheless the model
dependent value extracted from these data has a central value of \vubcb =0.079 
\cite{beauty96}, very close to the
exclusive result.

\begin{figure}[bt]
\vspace{-9mm}
\centerline{\epsfig{figure=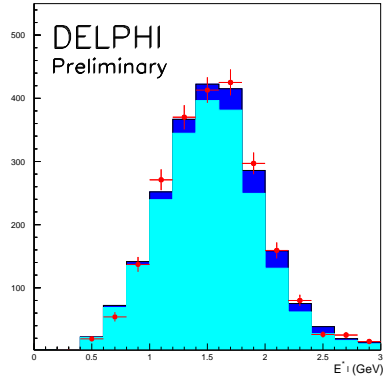,width=2.in}}
\vspace{-4mm}
\caption{\label{vubdelphi}The lepton energy distribution in the $B$ rest
frame from DELPHI. The data have been enriched in $b \to u$ events, and the 
mass of the recoiling hadronic system is required to be below 1.6 GeV. The 
points indicate data, the light shaded region, the fitted background and the 
dark shaded region, the fitted $b \to u \ell \nu$ signal. }
\end{figure}

Recently, interest has been stirred by a new approach 
\cite{vub-bdu} to the extraction of \vub\ 
based on the OPE
approach. The idea is that if the semileptonic width $\Gamma _u$ is extracted by 
integrating over the hadronic mass $X_u$ recoiling
against the lepton neutrino pair in a sufficiently large region of phase space, 
the relationship between \vub\ and the 
measured value of the charmless semileptonic branching fraction can be reliably 
predicted. All the LEP experiments
but OPAL attempted 
to use this technique to determine \vub . An example of the lepton spectrum 
obtained by cutting on the measured $X_u < 1.5$ GeV is shown in
Fig.~\ref{vubdelphi}. Note that the claimed \vub\ signal is accompanied by a 
dominant $b\dec c$ background. 
The three experiments combine their analyses and quote 
$|V_{ub}|= (4.05
^{+0.39}_{-0.46}(stat+det)^{+0.43}_{-0.51} (b\dec c) 
^{+0.23}_{-0.27}(b\dec u) \pm 0.02
(\tau _b) \pm 0.16 (HQE)\times 10^{-3}$ \cite{vubwg}. These results raise 
several concerns. First of all
the small error on the $b\dec c$ background component is not adequately 
justified. It implies a knowledge of
the $b\dec c$ background to better than 1\%. Moreover the validity of the 
theoretical input applies only
if the full phase space for the $b\dec u \ell \overline{\nu}$ is measured. Even 
if the mass cut $M_X<1.5$ GeV
should include most of the $\Gamma _u$ width, the complex set of event selection 
criteria 
applied may bias the phase space and make the errors in the \vub\ determination 
bigger than 
expected \cite{ligeti-dpf}. 

Much work needs to be done to achieve a precise measurement of \vub . This
is a quite important element of our strategy to pin down the CKM sector of the 
Standard Model.  
On the theoretical side, large efforts are put in developing more reliable 
methods to determine
the heavy to light form factors. A combination of several 
methods \cite{neub-buras}, all with a limited range of applicability,
seem to be the strategy more likely to succeed. For instance, lattice QCD can 
provide reliable estimates
of the form factors at large momentum transfer \cite{excl-lat}, where the 
discretization errors 
are under control.
HQET predicts a relationship between semileptonic $D$ decays and semileptonic 
$B$ decays. To check 
these predictions and apply them to \vub\ estimates, large data sample with 
reconstructed neutrino momentum 
are necessary. For now, I use the $\rho \ell \overline{\nu}$ result as the best 
estimate of \vubcb .
I take $|V_{ub}/V_{cb}|= 0.085 \pm 0.023$.

\subsection{The present knowledge of the CKM parameters}
In addition to the measurement of $|V_{ub}/V_{cb}|$, other experimental 
constraints can be used 
to check the  $bd$ unitarity triangle.  

\begin{figure}[htb]
%\vspace{-.8cm}
\centerline{\epsfig{figure=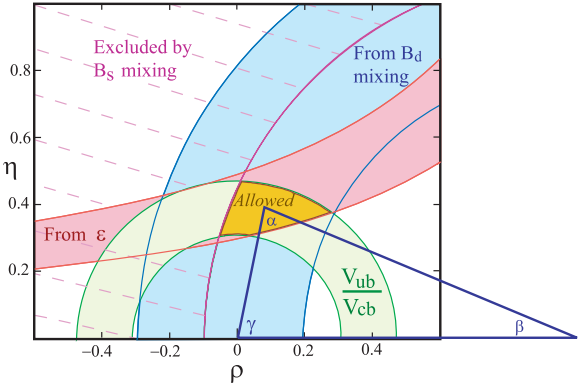,height=2in}}
\vspace{-0.2cm}
\caption{\label{stone-const}The CKM triangle 
shown in the $\rho-\eta$ plane. The
shaded regions show $\pm 1\sigma$ contours given by
$|V_{ub}/V_{cb}|$, neutral $B$ mixing, and CP violation in $K_L^o$ decay 
($\epsilon$) \cite{stone}. The dashed region 
is excluded by $B_s$ mixing limits.
The allowed region is defined by the overlap of
the 3 permitted areas, and is where the apex of the CKM triangle  sits.}
\end{figure}

An important input is the 
$\epsilon$ parameter 
describing 
$CP$ violation effects in $K$ decays \cite{kcp}.
Note that the $\epsilon$ parameter is the only constraint that 
forces $\eta$ 
to be positive. It is related to $\eta$ and $\rho$ through the relationship:
%\begin{equation}
\begin{displaymath}
\eta\left[(1-\rho)A^2(1.4\pm 0.2)+0.35\right]A^2{B_K \over 0.75}=\\
(0.30\pm 0.06),
\end{displaymath}
where the errors arise mostly from uncertainties on $|V_{cb}|$ and $B_K$. Recall 
that $A$ is
one of the Wolfenstein parameters:
\begin{equation}
A = V_{cb}/\lambda ^2,\\ |\lambda| =|V_{us}| =0.220\pm 0.002. 
\end{equation}
$B_K$
is taken as 0.75$\pm$0.15 according to Buras \cite{Buras_bk}.
Also the parameter $Re(\epsilon ^{\prime}/\epsilon )$, recently measured with 
increased
accuracy by KTeV \cite{ktev} and NA48 \cite{na48} can be used. Several
phenomenological calculations give $\eta$ as positive \cite{buras-eta}.
Note that a 
recent 
lattice calculation \cite{soni-epsilonp}, using a novel algorithm involving 
domain wall fermions (DWF),
obtains for the parameter $Re(\epsilon ^{\prime}_{DWF}/\epsilon _{exp})$, 
a negative value of $-(3.3\pm 0.3 _{stat}\pm 1.6_{sys})\times 10^{-2}\eta$,
where $\eta$ is the CP violation parameter in the Wolfenstein parameterization 
of the CKM matrix. 
This result is quite recent and needs further studies \cite{martinelli}, but it 
may imply a 
constraint in the $\rho - \eta$
plane not overlapping with the others and implies a violation of the Standard
Model.

Finally neutral $B$ meson oscillations provide additional constraints, 
semicircles in the $\rho -\eta$ plane 
centered at $(\rho,\eta)=(1,0)$.
The neutral $B_d$ oscillations 
are measured rather well \cite{lep-osc}. The world average, including data from 
LEP, CDF, and SLD, gives
$\Delta M_d=0.481 \pm 0.017 {\rm ps}^{-1}$. Only lower limits exist for $\Delta 
M_s$. The LEP average
is 9.6 ps$^{-1}$, dominated by the 1998 ALEPH inclusive 
lepton analysis corresponding to a 95\% C.L. lower limit of 9.5 ps$^{-1}$. The 
world average
presented at Tampere is 12.4 ps$^{-1}$, whereas the most recent preliminary 
world average is 14.3 ps$^{-1}$.
I will take as the world average the value of 12.4 ps$^{-1}$, because I believe 
that a preliminary
value that is so much bigger than any individual measurement needs further 
scrutiny.

The constraints on
$\rho$ versus $\eta$ from the $|V_{ub}/V_{cb}|$ determination, 
$\epsilon$ and $B$ 
mixing are shown in Fig.~\ref{stone-const}. The bands represent $\pm1\sigma$ errors,
for the measurements and a 95\% confidence level upper limit on $B_s$ mixing.
 The width of the $B_d$ mixing band 
is caused mainly by 
the uncertainty on $f_B$, taken here as $240> f_B > 160$ MeV. 

\begin{figure}[hbt]
\begin{center}
\epsfig{figure=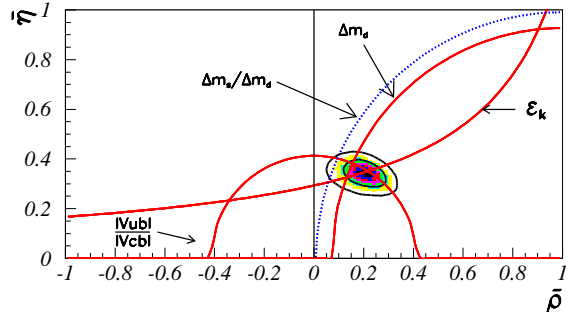,
bbllx=80pt,bburx=503pt,bblly=8pt,bbury=249pt,height=4cm}
\caption{ The allowed region for $\overline{\rho}$ and $\overline{\eta}$ 
according to Ref. \cite{sr}.
The contours at 68 $\%$ and 95 $\%$ are shown. The full lines correspond to the 
central values of the constraints given by
the measurements of  $  \frac{\left | V_{ub} \right |}{\left | V_{cb} \right |} 
$, $\epsilon$ and $\Delta m_d$.
The dotted curve corresponds to the 95 $\%$ C.L. upper limit obtained from the 
experimental limit on
$\Delta m_s$. The variables $\overline{\rho}$ and $\overline{\eta}$ are 
defined as
 $\overline{\rho}\equiv \rho(1-\lambda^2)$ and  $\overline{\eta}\equiv \eta(1-
\lambda^2)$. }
\label{fairytale}
\end{center}
\end{figure}
Some authors have taken the data discussed above as the input of a global fit 
that is supposed to generate
confidence level contours defined the Standard Model allowed region in the 
$\rho-\eta$ plane. Fig.~\ref{fairytale}
shows a recent example of such an analysis \cite{sr}. The big problem in works 
of this nature is how 
to combine in a reliable fashion results often dominated by theoretical 
uncertainties. Theoretical
errors are often educated guesses \cite{bigiwg} and thus not easily amenable to a 
rigorous statistical analysis.
The relevance of the small contour that defines the Standard Model allowed 
region in this plot is questionable.
For example, it will be very hard to attribute to a Standard Model failure any 
discrepancy between these
results and new data coming from the next generation $b$ experiments that will 
be discussed later.

\begin{figure}[hbt]
\begin{center}
\epsfig{figure=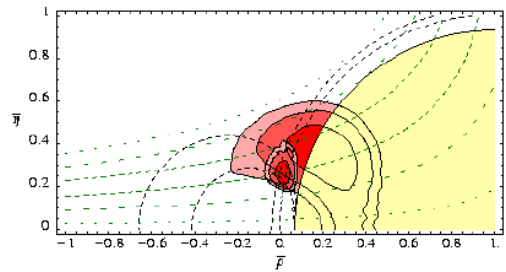,bbllx=80pt,bburx=550pt,bblly=400pt,
bbury=534pt,width=3in}
\vspace{2cm}
\caption{\label{bhrplot} Fit 1 (smaller region) from Ref. \cite{bhr}: prediction 
for $\rho$ and $\eta$ using the quark mass constraints
from Eq.~\ref{c1} and~ \ref{c2}. Fit. 2 (larger region) corresponds to a 
SM fit using \vubcb , \dms , \dmd\ but not $\epsilon$.
For both fits the countours are 68, 95 and 99 \% C.L. respectively. The 
variables $\overline{\rho}$ and $\overline{\eta}$ are defined as
 $\overline{\rho}\equiv \rho(1-\lambda^2)$ and  $\overline{\eta}\equiv \eta(1-
\lambda^2)$.}
\end{center}
\end{figure}

There is one interesting idea that was proposed originally by H. Fritzsch 
\cite{harald} and is explored in 
several recent papers. Its starting point is the observation that both the quark 
masses and the quark mixing
follow specific hierarchies. Perhaps these patterns contain a clue that can lead 
us to the more complete theory
that we are so eagerly pursuing. In particular quark mass textures,
i.e. specific patterns of zeros in quark mass matrices, that are motivated
by attempts to understand the origing of flavour assuming a spontaneously
broken symmetry, predict relationship between quark masses and quark mixing
parameters. For example, Barbieri, Hall and Romanino \cite{bhr} explored 
the constraints in the $\rho-\eta$ plane produced by the relationships:
\begin{eqnarray}
\left| \frac{V_{ub}}{V_{cb}} \right| =\left(\frac{m_u}{m_c}\right)^{1/2}\\ 
\label{c1}
\left| \frac{V_{td}}{V_{ts}} \right| =\left(\frac{m_d}{m_s}\right)^{1/2} 
\label{c2}
\end{eqnarray}
Fig.~\ref{bhrplot} shows the results from their analysis. Also in this case
the contour plots should be taken with some reservations. It should be noted 
that
their results are consistent with a Standard Model fit performed with
the same technique as Roudeau \etal . The consistency is modest. This analysis
illustrates another handle that we may use in the future more
effectively to reach a more complete understanding of the mystery of flavour.

\section{Rare $b$ Decays}

\begin{figure}[hbt]
%\vspace{1.0cm}
\centerline{\epsfig{figure=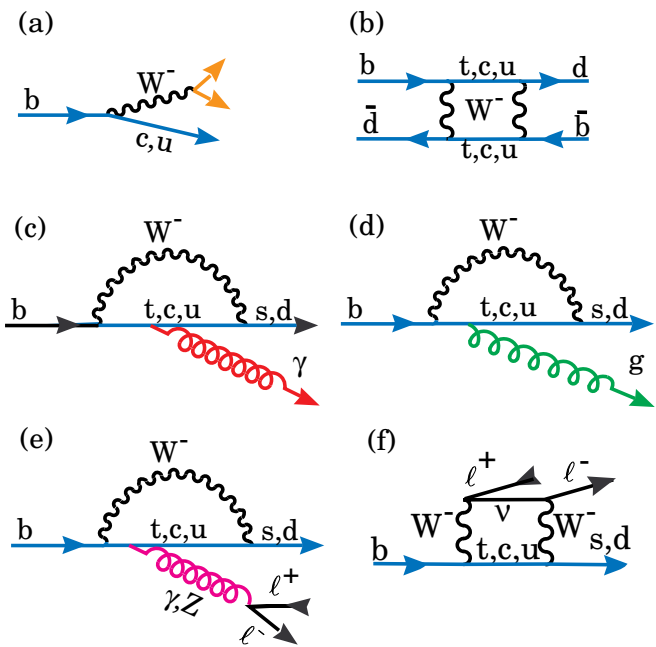,
height=5.0in,bbllx=0bp,bblly=0bp,bburx=600bp,bbury=800bp,clip=}}
\vspace{-2.5cm}
\caption{Feynman Diagrams for $b$ Decays}
\label{rareb}
\end{figure}

``Rare" $b$ decays encompass a large class of decay modes, with
branching fractions not necessarily exceedingly small, where a suppression
mechanism reduces the rate compared to the dominant ``tree diagram.''
Fig.~\ref{rareb} shows the Feynman diagrams for the various decay processes 
considered. Fig~\ref{rareb} (a) shows a tree amplitude that is associated
with $b\dec u$ CKM suppressed decays.

$B^0-\bar{B^0}$ 
mixing occurs via a 
``box" diagram with virtual $W$ bosons and $t$ quarks inside the box  
[Figure \ref{rareb}(b)]. 
The box diagram gives rise
to large fractions of mixed events: 17\% for $B^0$ and 50\% for $B_s$ mesons. 

Flavor-changing neutral currents lead to the transitions $b\to s$ and 
$b\to d$. 
These can be described in the Standard Model by one-loop diagrams, 
known as ``penguin" diagrams, where a $W^-$ is emitted and reabsorbed. 
The first such process to be observed was $b\to s\gamma$, described by the 
diagram in Figure \ref{rareb}(c), where the $\gamma$ can be radiated from any 
charged particle line. Another process which is important in rare $b$ decays 
is $b\to s g$, where $g$ designates a gluon radiated from a quark line
[Figure \ref{rareb}(d)]. A third example of such processes is the transition
$b\to s\ell^+\ell^-$ which can occur through the diagrams shown in Figures 
\ref{rareb}(e) and \ref{rareb}(f). We consider the loop processes shown in
 Figures \ref{rareb} (c)-(f)
to be among the most interesting and important rare $b$ decays.
 
\subsection{The decays $b\dec s \gamma$ and $b\dec s \ell ^+\ell ^-$}

There are several reasons why the study of rare $b$ decays is very important. 
First of all, the suppression involved in loop diagrams makes it possible for 
these 
Standard Model processes to interfere with decay diagrams mediated by exotic 
mechanisms due to `beyond Standard Model' interactions. In addition, loop 
diagrams and CKM suppression can affect our ability of measuring CP violation
asymmetries.

CLEO has recently reported an updated value of
this branching fraction, ${\cal B}(b\dec s \gamma)=(3.15\pm 0.35 \pm 0.32 \pm
0.26)\times 10^{-4}$ \cite{sgleppho}, where the first uncertainty is 
statistical, the second
is systematic and the third accounts for model dependence. 
ALEPH measured ${\cal B}(b\dec s \gamma)=(3.11\pm 0.8 \pm 0.72 \pm
0.26)\times 10^{-4}$ \cite{sgaleph}, where the first error is statistical and
the second is systematic, providing a nice confirmation of the CLEO
result. In parallel a new
theoretical estimate of the Standard Model prediction with a full
next-to-leading-order-log calculation \cite{n3} has been performed, giving
a predicted ${\cal B}(b\dec s \gamma)=(3.28\pm 0.33)\times 10^{-4}$, in
excellent agreement with the measured values.

The decay $b\dec s \gamma$ was the first inclusive rare $b$ decay to be 
measured \cite{bsgamma} and stirred a lot of theretical interest. It is 
a powerful  constraint of theories ``beyond the
Standard Model." For example, Masiero \cite{antonio-b96} identifies this 
process as ``the most relevant place in flavor changing neutral current $B$
physics to constrain SUSY, at least before the 
advent of $B$ factories.'' 

The Standard Model predicts no CP asymmetry in $b\dec s\gamma$. However recent
theoretical work \cite{n8} suggested that non-Standard Model physics may
induce a significant CP aysmmetry. Thus a search for CP asymmetry has been
performed by CLEO. A preliminary results of ${\cal A} = (0.16\pm 0.14\pm 0.05)
\times (1.0\pm 0.04)$ has been reported by CLEO \cite{sgleppho}. The 
first number is the central value,
followed by the statistical error and an additive systematic error. The 
multiplicative error is related to the uncertainty in the 
mistagging rate correction. From this result a 90\% C.L. limit on the CP
asymmetry ${\cal A}$ of $-0.09<{\cal A}<0.42$ is derived. These results are
based on $\approx 3.1\ {\rm fb}^{-1}$ of data. 

Another related decay mode that is quite important in constraining new
physics is the decay $b\dec s\ell ^+ \ell ^-$ and the related exclusive
channels $B\dec K^{(\star )} \ell ^+ \ell ^-$. These processes have been 
studied both at CLEO and at CDF and D0. Table~\ref{lpluslmin} summarizes
the searches for these decays.

\begin{table}
\begin{center}
\caption{\label{lpluslmin} Searches for $b\dec \ell ^+ \ell^-$ decays
compared with Standard Model theoretical predictions \cite{alill}.}
\begin{tabular}{lclc}
\br
$b$ decay & 90\% C.L. & Group & Prediction \\
mode & $\times 10^{-5}$ & ~~ &  $(\times 10 ^{-6})$ \\
\hline
$s\mu ^+ \mu^-$ & $32$ & D0 \cite{d02l} &  $(8\pm 2)$\\
~~ &5.7 &CLEO \cite{cleo2l} &~~\\
$K^{\star o}\mu ^+\mu ^-$ & 0.4 &CDF \cite{cdf2l}
& $2.9$ \\
~~ & $1.1$ &CLEO \cite{cleokse}&~~\\
$K^{\star o} e^+e^-$ & $1.3$ & CLEO \cite{cleokse} & 
$5.6$ \\
$K^-\mu ^+\mu  ^-$ & $0.83$ & CLEO \cite{cleokse} & $0.6$\\
~~ & $0.52$ & CDF \cite{cdf2l} & ~~\\
$K^-e^+e^-$ &$1.1$ & CLEO \cite{cleokse}&$0.6$\\
\br
\end{tabular}
\end{center}
\end{table}

\subsection{Rare Hadronic $b$ Decays}
Fig.~\ref{rareb} includes the dominant diagrams mediating 
rare $B$ hadronic decays.
The interplay between penguin and loop diagrams may affect our ability of
measuring CP asymmetries in two different ways. On one hand, loops and 
CKM suppressed diagrams 
can lead to final states accessible to both $B$ and $\overline{B}$ decays, 
making it possible to measure interference effects without neutral
$B$ mixing. On the
other hand, the interplay between these two processes can cloud the 
relationship between measured $CP$ asymmetries and the CKM phase when mixing
induced CP violation is looked for. A classical 
example 
of this effect is the decay $B^o\dec \pi^+\pi^-$. The main Standard Model 
diagrams contributing to this decay process are shown in Fig.~\ref{rareb} 
(a) and (d). 
If the $b\dec u$ diagram is dominant, the angle $\alpha$ can be extracted 
from the measurement of the asymmetry in the decay $B^o\dec \pi^+
\pi ^-$.  On the other hand, if these two diagrams 
have comparable amplitude, the extraction of $\alpha$ from this decay channel is
 going to be a much more difficult task. 

CLEO has studied several decays that can lead to a more precise understanding 
of the interplay between penguin diagrams and $b\dec u$ diagrams in $B$ meson
decays. The analysis technique used has been extensively refined in order to 
make the best use of the limited statistics presently available. In most of the
decay channels of interest, the dominant source of background are continuum
events; the fundamental difference
between $e^+e^-\dec q\overline{q}$ and $B$ decays is the shape of the underlying 
event. The
latter decays tend to produce a more `spherical' distribution of particles
whereas continuum events tend to be more `jet-like', with most of the particle
emitted into two narrow back to back `jets'. This property can be translated 
into
several different shape variables.
 CLEO constructs a Fisher discriminant 
${\cal F}=\Sigma _i \alpha_ iy_i$, a linear combination of several 
variables $y_i$. The variables used are $|\cos{\theta _{cand}}|$, the
cosine of the angle between the candidate $B$ sphericity axis and the beam
axis, the ratio of Fox-Wolfram moments $R_2=H_2/H_0$, and nine
variables that measure the scalar sum of the momenta of the tracks and showers
from the rest of the event in 9 angular bins, each of 10$^{\circ}$,
around the candidate sphericity axis. The coefficients $\alpha _i$ have
been chosen to optimize the separation between signal and background Monte 
Carlo samples \cite{rareb}. In 
addition, several kinematical constraints allow a more precise determination of
the final state. First of all  they consider the energy difference $\Delta 
E=E_{cand}-E_B$, where $E_{cand}$ is
the reconstructed  candidate mass  and $E_B$ is the known beam 
energy ($\Delta E =0$ for signal events) and the
beam constrained mass $M_B$ calculated from $M_B^2=E_{beam}^2-\left(\Sigma _i 
\vec{p}_i\right)^2$. In addition, the $B$ decay angle with 
respect to the beam axis has a $\sin ^2 (\theta _B )$ angular distribution. 
Finally,
to improve the separation between the final states $K\pi$ and $\pi\pi$ 
the specific energy loss in the drift chamber, $dE/dx$ is used. 

CLEO uses a
sophisticated unbinned maximum likelihood (ML) fit to optimize the precision of 
the
signal yield obtained in the analysis, using 
$\Delta E$, $M_B$, ${\cal F}$, $|\cos{\theta _B}|$, and $dE/dx$ wherever 
applicable. In each of these fits the likelihood of the event is parameterized
by the sum of probabilities for all the relevant signal and background 
hypotheses, with relevant weights determined by maximizing the likelihood 
function (${\cal L}$). The probability of a particular hypothesis is calculated
as the product of the probability density functions for each of the input 
variables determined with high statistics Monte Carlo samples.
The likelihood function is defined as:
\begin{equation}
{\cal L} = \Pi _k \Sigma _i P^i_k(\Delta E, M_B,{\cal F}, dE/dx, 
\cos{\theta _B}) \times f_i
\end{equation}
where the index $k$ runs over the number of events, the index $i$ over
the hypotheses considered, $P^i_k$ are the probabilities for
different hypotheses obtained from Monte Carlo
simulations of the signal and background channels considered and 
independent data samples, and $f_i$ are 
the fractional yields for 
 hypothesis $i$, with the constraint:
\begin{equation}
\Sigma _i f_i =1
\end{equation}

Further details about the likelihood fit can be found elsewhere \cite{rareb}.
 The fits
include all the decay channels having a similar topology. For example, in the
final state including two charged hadrons, the final states considered were
$K^{\pm}\pi^{\mp }$, $\pi^+\pi^-$ and $K^+K^-$.

\begin{figure}[htb]
\centering
\leavevmode
\epsfxsize=3.0in
\epsffile{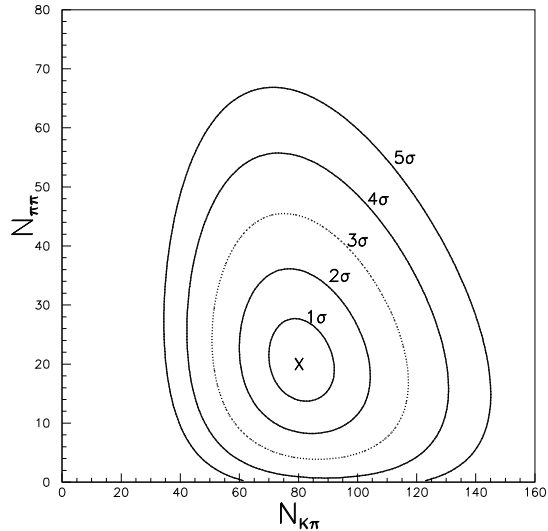}
\vspace{-1in}
\caption{Contours of the $-2\ln{\cal L}$ for the ML fit to 
$N_{K^{\pm}\pi^{\mp}}$ and $N_{\pi^+\pi^-}$, the $K^\pm\pi^\mp$
and $\pi^+\pi^-$ yields respectively. }
\label{fig:contourkpi}
\end{figure}

\begin{figure}[htb]
\centering
\leavevmode
\epsfxsize=3in
\epsffile{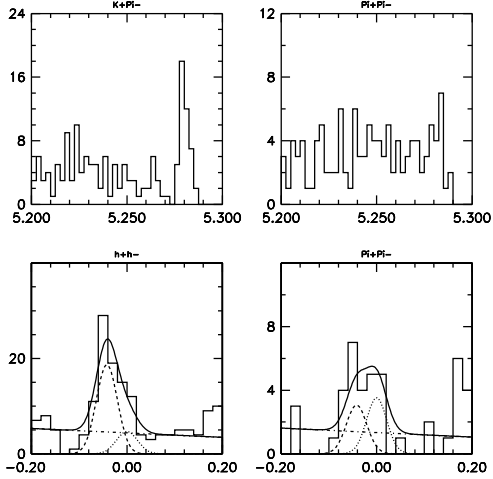}
%\vspace{-0.1cm}
\caption{Projections of $K\pi$ and $\pi\pi$ events
onto $M$ and $\Delta E$ with cuts.
Upper left: $M$ distribution of $K\pi$-like events;
upper right: $M$ distribution of $\pi\pi$-like events.
Lower left: $\Delta E$ distribution of events prior
to $\pi\pi$ vs $K\pi$ vs $KK$ selection according to dE/dx;
Lower right: $\Delta E$ distribution of events that are
more likely to be $\pi\pi$ than $K\pi$ or $KK$ based on dE/dx.
Overlays in the lower plots are the results of the likelihood
fit scaled by the efficiency of the cuts used to project
into these plots. Solid line: total fit; dashed: $K\pi$;
dotted: $\pi\pi$; dot-dash: continuum background.}
\label{fig:kpi-pipi-projections}
\end{figure}

Fig. \ref{fig:contourkpi} shows contour plots of the ML fits for the signal 
yields
in the  $K^{\pm}\pi^{\mp}$ and $\pi^+\pi^-$ final states. The other
channels included in the likelihood function have $f_i$ fixed to their most
probable value extracted from the fit. It can 
be seen that there is a well defined signal for both the final states  
$K^{\pm}\pi^{\mp}$ and
$\pi^+\pi^-$. The ratio between the most probable yields in
the two channels shows that the $b\dec u$ diagram
is suppressed with respect to the penguin diagram in $B$ decays to two 
pseudoscalar mesons.  Fig. \ref{fig:kpi-pipi-projections}
shows distributions in $M_B$ and $\Delta E$ for events after cuts on the
Fisher discriminant and whichever of $M$ and $\Delta E$ is not being
plotted, plus an exclusive classification into $K\pi$-like and $\pi\pi$-like
candidates
based on the most probable assignment with $dE/dx$ information. The
likelihood fit, suitably scaled to account for the efficiencies of the
additional cuts, 
is overlaid in the $\Delta E$ distributions to illustrate the separation
between $K\pi$ and $\pi\pi$ events.

Table \ref{twops} summarizes the CLEO results for the 
$B\dec K\pi,~\pi \pi,~ KK$ final states. Unless explicitly stated, the 
preliminary results are
based on 9.66 million $B\overline{B}$~pairs collected with the CLEO
detector. 

There is a consistent 
pattern of penguin dominance in \B\ decays into two charmless 
pseudoscalar
mesons that makes the prospects of extracting the angle  
$\alpha$ from the study of the CP asymmetry in
the $B\dec \pi^+\pi^-$ mode less favorable than originally expected.

\begin{table}
\begin{center}
\caption{Experimental results. 
Branching fractions
(${\cal B}$) and 90\% C.L. upper limits are given in units
of $10^{-6}$.}
\begin {tabular}{lll}

\br
Mode& 
${\cal B}$($10^{-6}$) & Signif.($\sigma$) \\
\hline
$\pi^+\pi^-$ &
   $4.7^{+1.8}_{-1.5}\pm 0.6$ &4.2      \\

$\pi^+\pi^0$ 
 
&$<12$  & 3.2        \\
\br
$K^+\pi^-$   
 & $18.8^{+2.8}_{-2.6} \pm 1.3$  &11.7      \\
$K^+\pi^0$  
& $12.1^{+3.0+2.1}_{-2.8-1.4}$  & 6.1        \\
$K^0\pi^+$  & 
$18.2^{+4.6}_{-4.0} \pm 1.6$   
 & 7.6 \\
$K^0\pi^0$   
&$14.8^{+5.9+2.4}_{-5.1-3.3}$    & 4.7            \\
\br
$K^+K^-$     
& $<2.0$ & --  \\
$K^+\overline{K}^0$   &
 $<5.1$   & 1.1  \\
\br
\end {tabular}
\label{twops}
\end{center}
\end {table}

\begin{figure}[htb]
%\vspace{1.0cm}
\centerline{\epsfig{figure=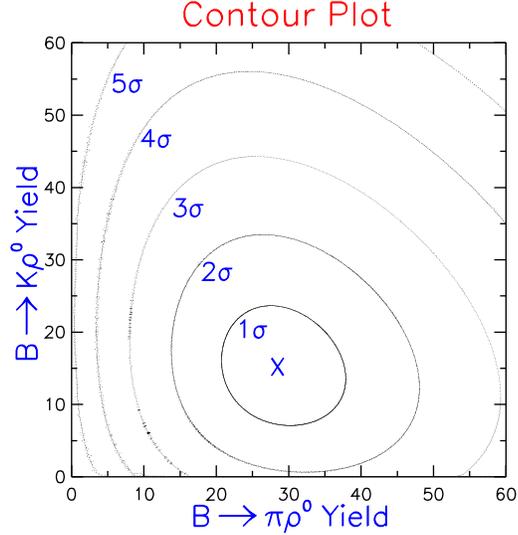, width=2.8 in}}
%\vspace{1cm}
\caption{$n\sigma$ contours for the decays $B^+ \dec \rho ^0 \pi^+$
and $B^+\dec \rho ^0 K^+$. The cross identifies the most 
probable values of the yields
for these two channels.
\label{pirho}}
\end{figure}

 Recent
CLEO data suggest that final states involving a vector and a pseudoscalar meson 
offer a different picture. In fact,   
$B^+ \dec \pi ^+ \rho^o$ and $B \dec \rho ^{\pm} \pi ^{\mp}$ \cite{gao}
have been observed, while the penguin dominated final states $B^+ \dec K  \rho$ 
and $B \dec K^{\star} \pi$ have
not. 
 The analysis procedure is similar to the one
adopted for the charmless pseudoscalar-pseudoscalar exclusive decays. In this
quasi two-body case, there are three 
particles in the final state and some additional constraints are provided
by the vector particle decay kinematics. The invariant mass of its decay
products must be consistent with the vector meson mass. In addition, the
vector meson is polarized, thus its helicity angle
$\theta _h$ has a $\cos ^2{\theta _h}$ distribution. The 
maximum likelihood fit
includes $\pi\rho^0$ and $K\rho ^0$ signal channels and continuum samples. The 
$n\sigma$ contour
 plot for this analysis is shown 
in Fig. \ref{pirho} and gives solid evidence for a $B^+ \dec \pi ^+\rho ^o$ 
signal, while the $B^+\dec K^ +\rho ^o$ channel appears to be suppressed.
Table~\ref{results2} shows the results for each decay mode
 investigated. For observations the final
 result is reported as a branching fraction central value, while for modes
 where the yield is not sufficiently significant the 90\% confidence
 level upper limit is quoted. Also listed in the Table are 
theoretical
 estimates.

\begin{table*}[htb]
\caption{ Summary of CLEO results on charmless hadronic
decays to a final state including a vector
and a pseudoscalar meson compared with expectations from theoretical models.}
\vspace{0.4cm}
\begin{center}
\begin{tabular}{llll}
\br
Decay mode & $\calB (10^{-6})$  & Theory \calB\ ($10^{-6}$)&References\\
\hline
\Bomegapi   & $11.3^{+3.3}_{-2.9}\pm{1.5}$ & $0.6-11$        &
    \cite{chau,dean,kps,du,oh,ciuchini,ali,cheng}     \\
\Bomegapiz  &          $< 5.8 $            & $0.01-12$       &
    \cite{chau,dean,du,oh,ciuchini,ali,cheng}         \\
\Bomegak    &          $< 8.0$             & $0.2-13$        &
    \cite{chau,dean,kps,du,oh,ciuchini,ali,cheng}     \\
\Bomegakz   &          $< 21$               & $0.02-10$       &
    \cite{chau,dean,du,oh,ciuchini,ali,cheng}         \\
\Bomegah    & $14.3^{+3.6}_{-3.2}\pm 2.1$  &                 & 
                        \\
\Bomegarhop &          $< 47$               & $7-28$          &
    \cite{chau,dean,kpsvv,ciuchini,ali,cheng}         \\
\Bomegarhoz &          $< 11$               & $0.005-0.4$     &
    \cite{chau,ciuchini,ali,cheng}                    \\
\Bomegakstp &          $< 52$              & $0.9-15$        &
    \cite{chau,dean,kpsvv,ciuchini,ali,cheng}         \\
\Bomegakstz &          $< 19$              & $0.3-12$        &
    \cite{chau,dean,ciuchini,ali,cheng}               \\
\Brhozpi    &    $15^{+5}_{-5} \pm 4$       & $0.4-8$      &
  \cite{chau,ebert,dean,kps,du,oh,ciuchini,ali,cheng} \\
\Brhompi    &    $35^{+11}_{-10} \pm 5$    & $26-52$      &
  \cite{chau,ebert,dean,du,oh,ciuchini,ali,cheng}     \\
\Brhozpiz   &          $<5.1$              & $0.9-2.3$       &
  \cite{chau,ebert,du,oh,ciuchini,ali,cheng}          \\
\Brhozk     &          $<22 $              & $0.1-1.7$    &
  \cite{desh,chau,dean,kps,du,oh,ciuchini,ali,cheng}  \\
\Brhomk     &          $<25 $              & $0.2-2.5$    &
  \cite{desh,chau,dean,du,oh,ciuchini,ali,cheng}      \\
\Brhozkz    &          $<27 $              & $0.04-1.7$   &
  \cite{desh,chau,dean,du,oh,ciuchini,ali,cheng}      \\
\Bkstzpi    &          $<27 $              & $4-12$       &
  \cite{desh,chau,fl,du,oh,ciuchini,ali,cheng}        \\
\Bkstppi    & $22^{+8 \; +4}_{-6 \; -5}$   & $1.2-19$     &
  \cite{desh,chau,dean,du,oh,ciuchini,ali,cheng}      \\
\Bkstzpiz   &          $<4.2$              & $1.1-5$         &
  \cite{desh,chau,dean,du,oh,ciuchini,ali,cheng}      \\
\Bkstpk     &          $<6  $              &              &
                                                      \\
\Bkstzk     &          $<12 $              & $0.2-1$      &
  \cite{chau,kps,du,oh,ali,cheng} \\
\Bphipi     &          $<4.0$              & $0.001-0.4$  &
    \cite{xing,dean,fl,kps,du,oh,ciuchini,ali,cheng}  \\
\Bphipiz    &          $<5.4$              & $0.0004-0.2$ &
    \cite{xing,dean,fl,du,oh,ciuchini,ali,cheng}      \\
\Bphik      &          $< 5.9$             & $0.3-18$     &
    \cite{desh,chau,dean,fl,dav,kps,du,oh,ciuchini,ali,cheng} \\ 
\Bphikz     &          $< 28$              & $0.3-18$     &
    \cite{desh,chau,dean,fl,dav,du,oh,ciuchini,ali,cheng}     \\ 
\br
\end{tabular}
\end{center}
\label{results2}
\end{table*}

\begin{figure}[htb]
\vspace{-0.4cm}
\centerline{\epsfig{figure=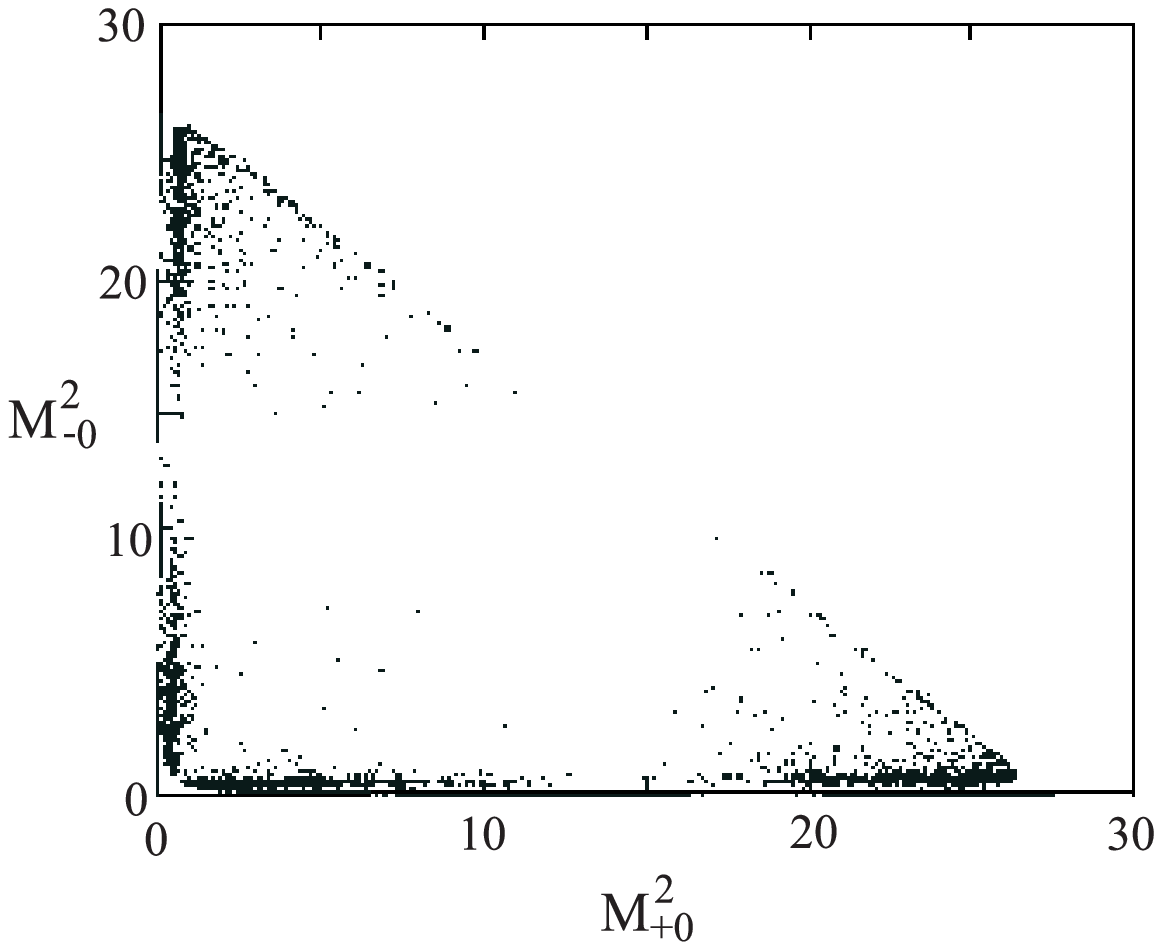,height=2.3in}}
\vspace{-.6cm}
\caption{\label{dalitz} The Dalitz plot for
$B^o\to\rho\pi\to\pi^+\pi^-\pi^o$ from Snyder and Quinn.}
\end{figure}

The stronger $b\dec u$ component in vector to pseudoscalar final states suggests 
that
an approach, proposed by Snyder and Quinn \cite{sq} can lead to
a precise measurement of the angle $\alpha$.
The interference between Tree and Penguin diagrams can be measured
 through the time dependent CP violating
 effects in the decays $B^o\to\rho\pi\to\pi^+\pi^-\pi^o$. 
 Since the 
$\rho$ is spin-1, the $\pi$ spin-0 and the initial $B$ also spinless, the 
$\rho$ 
is fully polarized in the (1,0) configuration, so the helicity angle has a 
$\cos^2{\theta}$
distribution.
 This causes the periphery of the Dalitz plot to be 
heavily populated, especially the corners. A sample Dalitz plot is shown in 
Fig.~\ref{dalitz}. 
 Snyder and Quinn performed an idealized analysis that showed that a background 
free 1000 or
2000 flavor tagged event samples is sufficient to measure $\alpha$. 

These charmless hadronic decays can, in principle, provide also
some information on the phase $\gamma$, conventionally defined as
the argument of $V_{ub}$. This phase can cause asymmetries in decay rates to 
charge conjugate pairs. CLEO has taken a first look at possible evidence for 
these rate asymmetries. No positive signal has been seen and the present 
sensitivity
is shown in Table~\ref{tbl:sum} summarizing their preliminary results.
\vskip 0.5cm
\begin{table*}[ht]
\begin{center}
\caption{\acp\ measurements in five charmless $B$ decay modes. CLEO 1999 
Preliminary.}
\begin{tabular}{lllll}
\br
%\noalign{\hrule\vskip 0.15truecm}
Mode &  $\sig$              & $\sigbar$            & $\acp$ & 90\% CL          
\cr
\hline
$K\pi^0$         &  $16.8\pm 7.5$      & $28.9\pm 7.5$        & $-0.27\pm 
0.23\pm 0.05$          & $[-0.70, 0.16]$   \cr
$K^0_S\pi$       & $14.5 \pm 4.4$      & $10.2 \pm 4.0$       & $0.17\pm 0.24\pm 
0.05$           & $[-0.27, 0.61]$   \cr
$K\pi$           &$38.6_{-8.1}^{+9.0}$ &$41.6_{-8.0}^{+8.9}$  & $-0.04\pm 
0.16\pm 0.05$          & $[-0.35, 0.27]$   \cr
$\etapr K$       & $51.7\pm 9.2$       & $48.7\pm 8.9$        & $0.03\pm 0.12\pm 
0.05$           & $[-0.22, 0.28]$  \cr
$\omega\pi$     &$\aerr{9.4}{4.9}{4.0}$&$\aerr{19.1}{6.8}{5.9}$& $-0.34\pm 
0.25\pm 0.05$         & $[-0.80, 0.12]$  \cr
\hline
\end{tabular}
\label{tbl:sum}
\end{center}
\end{table*}

\begin{table*}[htb]
\caption{Summary of CLEO results for \B\ branching fractions
to charmless final states involving an $\eta$ or $\eta^{\prime}$
mesons. The second  column gives recent
preliminary result \cite{leppho}, unless otherwise stated. The central value 
followed by statistical and
systematic error is given if significantly above zero, otherwise as a 90\%
confidence level upper limit is quoted.  Estimates from various theoretical
sources are shown for comparison.}
%\medskip
%\hbox{\hspace{-1.5cm}
%\def\notext{ & & & & \cr}
\begin{center}
\begin{tabular}{llll}
%**%\dbline
\br
Decay mode 
&  $\calB (10^{-6})$ & Signif. ($\sigma$) &
 Theory \calB($10^{-6}$)\cr
\hline
$B^+\goto\etaprkp$     &  \retapKp &
                        16.8 &  
                        7 -- 65 \cite{chau,du}\\
$B^0\goto\etaprkz$    & \retapKz & 11.7  &
                       9 -- 59 \cite{chau,du}\\
$B^+\goto\etaprpi$    & $<11$ & $0.2$ &
                        1 -- 23 \cite{chau,du}\cr
$B^0\goto\etaprpiz$\cite{etapprl}   & $<11$ & --
                         & 0.1 -- 14 \cite{chau,du}\cr
$B^+\goto\etaprkstp$\cite{cleoDPF}  &$<87$ &   1.2 & 
                          0.1 -- 3.7 \cite{chau,du}\cr
$B^0\goto\etaprkstz$\cite{cleoDPF}  & $<20$   &1.0 &
                         0.1 -- 8.0 \cite{chau,du}\cr
$B^+\goto\etaprrhop$ \cite{etapprl}  & $<47$ & -- 
                        & 3 -- 24 \cite{chau,du}\cr
$B^0\goto\etaprrhoz$   \cite{etapprl}&  $<23$ & --
                       & 0.1 -- 11 \cite{chau,du}\cr
$B^+\goto\etak$      & $<7.1$   & 1.0 &
                   0.2 -- 5.0 \cite{chau,du}\cr
$B^0\goto\etakz$      & $<9.5$ &--&
                        0.1 -- 3.0 \cite{chau,dean,du}\cr
$B^+\goto\etapi$      & $<6.0$ & 0.6 &
                       1.9 -- 7.4 \cite{chau,dean,du}\cr
$B^0\goto\etapiz$      & $<3.1$ & -- &
                        0.2 -- 4.3 \cite{chau,du}\cr
$B^+\goto\etakstp$   & \retaKstp &   4.8 &
                       0.2 -- 8.2 \cite{chau,du}\cr
$B^0\goto\etakstz$    &\retaKstz &  5.1 & 
                        0.1 -- 8.9 \cite{chau,dean,du}\cr
$B^+\goto\etarhop$  & $<16$ &  1.3 &
                       4 -- 17 \cite{chau,dean,du}\cr
$B^0\goto\etarhoz$   & $<11$  & 1.3 
                       & 0.1 -- 6.5 \cite{chau,dean,du}\cr
\br
\end{tabular}
\end{center}
\label{combtab}
\end{table*}
  
The branching fractions themselves can be used to constraint the angle 
$\gamma$. Several approaches have been pursued and they are
more or less model dependent \cite{gamma-lit}. Atwood and Soni 
performed a general analysis 
of $B$ decays involving 10 $B$ decay modes to pseudoscalar final states 
\cite{as98} for the charged
and neutral \B\ mesons respectively. Their
analysis has been recently updated to include the results discussed in this 
paper.  Conjugate
pairs have been taken together since the experimental data are not yet 
sensitive to CP violation effects. Rescattering effects are attributed to a 
single complex 
parameter $\kappa$. This analysis determines 8 parameters,  
tree amplitude $t$, color suppressed
amplitude $\hat{t}$, penguin amplitude $p$, electroweak penguin amplitude
$e$, real $\kappa =\kappa _R$, imaginary  $\kappa =\kappa _I$, and the 
CKM parameters $\rho$ and $\eta$. A self-consistent solution
is sought for using a $\chi ^2$ minimization. Several fits are performed 
using different simplifying assumptions. The best fits are obtained 
assuming that electroweak penguin and rescattering effects are negligible. In 
this case the authors obtain a value of
$\gamma = 92\pm 20$ degrees. 

Recently Hou, Smith and W\"uerthwein \cite{hsw} have performed a
similar analysis, including final states
 composed by two pseudoscalar mesons and a vector and a pseudoscalar
 meson. They perform a fit with the same procedure
as one of the fits by Atwood and Soni, with the restrictive assumption of 
factorization. They obtain $\gamma = 114 ^{+25}_{-21}$ degrees. The 
$VP$ final states have been studied with flavor SU(3) by Gronau
and Rosner \cite{gr99} and they found that several processes are consistent 
with $\cos{\gamma}<0$.  These results are preliminary.
These studies illustrate the increasing interest in these decays to measure 
fundamental Standard Model parameters. However to reach definite 
conclusion, a deeper understanding of hadronic physics must be achieved.

The CLEO study of $B$ decays to final states including two charmless hadrons
has presented some other interesting surprises. A large branching ratio
has been discovered for final states including a $\eta ^{\prime}$ meson. The
analysis technique in this case is essentially the same as the one discussed
above. The $\eta ^{\prime}$ is reconstructed both in its
$\eta \pi^+\pi^-$ and $\rho \gamma$ decay channels. The results for different 
decay modes including 
$\eta$ and $\eta ^{\prime}$ 
are summarized in Table~\ref{combtab}. The most notable 
feature is the astonishing large rate for the 
$\eta ^{\prime} K$ 
final state. While several theoretical interpretations have
been proposed to explain this enhancement \cite{etap-th}, a very
plausible explanation  still points to a dominance of penguin effects.

\section{The lesson from charm}
Charm is a unique probe of the Standard Model in the up quark sector, 
thus providing a window of opportunity to discover new physics complementary
to that attainable from the down-quark sector. Due to the effectiveness of 
the GIM mechanism, short distance contribution to
rare charm processes
are extremely small. Thus very often long range effects, for which reliable 
calculations are rarely available, dominate the rates. On the other hand the
fact that the Standard Model predictions are so small for flavor changing 
neutral currents in the charm sector provides a unique opportunity
to discover new physics in charm rare decays, $D^o\bar{D}^o$ mixing and 
CP violation in the decays of charmed mesons. 
\subsection{Rare and Forbidden decays of charmed mesons}
Flavor changing neutral current (FCNC) decays of the $D$ meson include the 
processes $D^o\dec \ell ^+\ell ^-,\ \gamma\gamma$, and $D\dec X_u \gamma,\ 
X_u\nu\overline{\nu},\ X_u \ell ^+\ell ^-$. 
They proceed via electromagnetic or weak penguin diagrams, with contributions 
from box diagrams in
some cases. Calculations of short distance contributions to these decays are 
quite reliable \cite{joanne96}, however the long range contribution estimates 
are plagued with hadronic uncertainties and are mentioned here 
to give an approximate upper limit of the size of these effects.

Table~\ref{rared} summarizes some theoretical predictions together with the 
experimental upper limits 
coming from fixed target charm experiments, most notably E791 \cite{e791r},
 and CLEO.  Soon another fixed target experiment at Fermilab, FOCUS, should 
provide an improvement of about a factor of 10 \cite{daniele}. These decays 
are usually classified into three categories:
\begin{itemize}
\item[1)] Flavor changing neutral current (FCNC) decays, such as 
$D^o\dec \ell ^+\ell^-$ and $D^o\dec X\ell ^+\ell ^-$,
\item[2] Lepton family number violating (LFNV) such as 
$D^+\dec h^-\mu^{\pm} e^{\mp})$,
\item[3)]Lepton number violating (LNV) decays such as 
$D^+\dec h^-\mu ^+\mu^+$.
\end{itemize}

The first decays (FCNC) are rare, namely they are processes that can proceed
via an internal quark loop in the Standard Model. The other two categories
are
strictly forbidden within the Standard Model. However they are are allowed  
in some of its extensions, such as extended technicolor \cite{joanne96}. The 
expected rates for class (1) decays
 are several orders of magnitude below
the experimental sensitivity. However this may represent a window of 
opportunity to discover new physics.

\begin{table*}[hbt]
\begin{center}
\caption{\label{rared}Measured experimental limits for various rare $D$ 
meson decays together with the  Standard Model predictions for the branching 
fractions due to short (${\cal B}_{SD}$) and long distance
 contributions ((${\cal B}_{LD}$).}
\begin{tabular}{llll}
\br
Decay Mode & Experimental limit (90 \% C.L.) &
${\cal B}_{SD}$) & (${\cal B}_{LD}$) \\  
\hline
$D^o\dec \mu^+\mu^-$ & $4.1\times 10^{-6}$ \cite{pdg98}
 & $(1-20)\times 10^{-19}$ &$<3\times 10^{-15}$\\
$D^o\dec e^+e^-$ & $6.2\times 10^{-6}$ \cite{e791r}
$(2.3-4.7)\times 10^{-24}$ & -- \\
$D^o\dec \mu^{\pm}e^{\mp}$ & $8.1\times 10^{-6}$ \cite{e791r}& $0$ &$0$\\
\hline
$D^o\dec \gamma\gamma $& -- & $10^{-16}$ & $<10^{-9}$ \\
\hline
$D\dec X_u\gamma$ & ~~& $1.4 \times 10^{-7}$ & -- \\
$D^o\dec \rho ^o\gamma$ & $2.4\times 10^{-4}$ \cite{pdg98}
& -- & $<2\times 10^{-5}$\\
$D^o\dec \phi \gamma$ & $1.9\times 10^{-4}$  \cite{pdg98}
& -- & $<10^{-4}$ \\
$D^+\dec\rho ^+\gamma$ & ~~ & ~~ & $2\times 10^{-4}$ \\
\hline
$D^o\dec X_u\ell ^+ \ell ^-$ & ~~ & $4\times 10 ^{-9}$ & ~~ \\
$D^o\dec \pi ^o ee/\mu\mu$ &$(4.5/2.6)\times 10^{-5}$  \cite{pdg98}
& ~~ & ~~ \\
$D^o\dec \overline{K}^o ee/\mu\mu$ &$( 1.1/2.6)\times 10^{-4}$  \cite{pdg98}
& ~~ & $<2\times 10^{-15}$ \\
$D^o\dec \rho ^o ee/\mu\mu$ & $1.0/2.3\times 10^{-4}$  \cite{pdg98}
&~~ & ~~ \\
$D^+\dec \pi^+ ee/\mu\mu $ & $(5.2/1.5)\times 10^{-5}$ \cite{e791r} & 
few$\times 10^{-10}$ & $<10^{-8}$ \\
$D^+\dec K^+ ee/\mu\mu $ & $(20/4.4)\times 10^{-5}$ \cite{e791r}& 
few$\times 10^{-10}$ & $<10^{-8}$ \\
\br
\end{tabular}
\end{center}
\end{table*}

\subsection{CP Violation and $D^o\overline{D}^o$ Mixing}

Another window of opportunity towards new physics discoveries is provided
by the study of CP violation and flavour oscillations in charm decays. Both
effects are expected to be highly suppressed. CP violation asymmetries are
expected to be around $10^{-3}$ \cite{bucella} in Cabibbo suppressed decay 
such as $D^+\dec K^-K^+\pi ^+$ and $D^o\dec K^-K^+$. Recent preliminary
results from Focus \cite{daniele} based on 59\% of their data have given
${\cal A}_{CP} = -0.004\pm 0.017$ for the decay mode $D^+\dec K^-K^+\pi ^+$ 
and ${\cal A}_{CP}=0.003\pm 0.039$ for $D^o\dec K^-K^+$. 

In the case of the $D^o$ the asymmetry may be a combination of direct
CP asymmetry and indirect CP asymmetry mediated by mixing. The 
$D^o\overline{D}^o$ mixing parameter $\Delta M_D$ is affected by small
distance contributions described by box diagrams via internal $d,\ s,\ b$ 
quarks and long distance contributions that are more difficult to evaluate. 
At any rate, the Standard Model expectations for $\Delta M_D$ are between 
$10^{-18}$ and $10^{-16}$ \cite{joanne96}.

The probability that a $D^o$ oscillates into a $\overline{D}^o$ is generally
expressed as:
\begin{equation}
I(t) \propto \left|\frac{q}{p}\right|^2 e^{-\Gamma _1 t}\left[1+e^{\Delta \Gamma t}+2 e^{0.5\Delta \Gamma t} \cos{\Delta mt}\right]
\end{equation}
where $\Delta m\equiv m_1 -m_2$, $\Delta \Gamma = \Gamma _1 -\Gamma _2$ and
the
subscripts 1 and 2 identify the two neutral $D$ meson mass eigenstates
and the complex parameters $p$ and $q$ account for CP violation  effects. 

The parameter $r_{WS}$ for the decay $D^o\dec K^+\pi ^-$ can be expressed as:

\begin{equation}
r_{ws}\approx e^{-\Gamma t} \begin{array}{l}[ r_{D} +r_{mix}
\times \frac{\Gamma ^2 t^2}{2} + \\
 \sqrt{2r_{mix}r_{D}}\cos{\Phi}\times \Gamma t]
 \end{array}
\end{equation}
where the first term is the $DCSD$ contribution parameterized by the quantity 
$r_D$, the second is the one due to mixing and the last one is an 
interference term between mixing and $DCSD$. The latter term
is set to zero for simplicity in some of the experimental estimates of 
$r_{mix}$, but this is not a legitimate assumption. Alternatively, we can parameterize $r_{ws}$ as:
\begin{equation}
r_{ws}= e^{-t}\begin{array}{l}[R_D+\sqrt{R_D}y^{\prime }t+ \\
+ \frac{1}{4}({x^{\prime}}^2 + {y^{\prime}}^2)t^2]
\end{array}
\label{dmix2}
\end{equation}
where $R_D$ is the rate of the direct doubly-Cabibbo-suppressed decay
$\overline{D}^o\dec K^+\pi ^-$, $x^{\prime}=x\cos{\delta} +y \sin{\delta}$
 and $y^{\prime} = y\cos{\delta} - x\sin{\delta}$ and $x =\Delta
M/\overline{\Gamma}$, $y=\Delta \Gamma/\overline{ \Gamma}$. $\overline
{\Gamma}$ represents the average lifetime and $\delta$
represents
a strong phase between the doubly-Cabibbo-suppressed decay and the Cabibbo
favored one. 
CLEO \cite{cleodmix} has recently used the formulation in Eq. \ref{dmix2} to obtain a very
precise measurement of the parameter $R_D=0.50^{+0.11}_{-0.12}\pm 0.08$. Their
limit at 95 \% C.L. are $(1/2) {x^{\prime}}^2<0.05\%$ and $-5.9\%<y^{\prime}
<0.3\%$.  E791 has extracted the 
mixing parameter $r_{mix}$ \cite{rmixsl} also from the semileptonic decay 
$D^o\dec K\ell \nu$. This leads to a more straightforward determination
because the $DCSD$ contribution is absent, obtaining $r_{mix}<0.50\%$
at 90\% C.L. Thus the amplitudes that describe $D^o\overline{D}^o$ are still
consistent with 0. Soon CLEO and FOCUS results based on 
semileptonic data will be available \cite{daniele}.

\section{Outlook}
The selected sample of flavour physics topics that I have discussed shows
that
there are several predictions of the Standard Model that we have yet to test 
thoroughly. In particular we must explore the phenomenology of CP
violation in $b$ and $c$ decays, flavour oscillations in the $B_s$ system and 
study rare decays of beauty and charm. Our ability to challenge the Standard 
Model and to discover new physics will rely on two different areas where 
progress is needed. On one hand, we need to reach a new level of 
sophistication in our experimental study of these decays. On the other hand, 
we need to refine the theoretical inputs, to develop less model dependent 
predictions for a larger class of branching fractions and CP asymmetries.

\begin{table*}[hbt]
\begin{center}
\caption{Required Measurements for $b$'s}
\label{cp}
\vspace*{2mm}
\begin{tabular}{llcccc} \hline\hline
Physics & Decay Mode & Hadron & $K\pi$ & $\gamma$ & Decay \\
Quantity&            & Trigger & sep   & det & time $\sigma$ \\
\hline
$\sin(2\alpha)$ & $B^o\to\rho\pi\to\pi^+\pi^-\pi^o$ & $\surd$ & $\surd$& $\surd$ 
&\\
$\cos(2\alpha)$ & $B^o\to\rho\pi\to\pi^+\pi^-\pi^o$ & $\surd$ & $\surd$& 
$\surd$ &\\
sign$(\sin(2\alpha))$ & $B^o\to\rho\pi$ \& $B^o\to\pi^+\pi^-$ & 
$\surd$ & $\surd$ & $\surd$ & \\
$\sin(\gamma)$ & $B_s\to D_s^{\pm}K^{\mp}$ & $\surd$ & $\surd$ & & $\surd$\\
$\sin(\gamma)$ & $B^-\to \overline{D}^{0}K^{-}$ & $\surd$ & $\surd$ & & \\
$\sin(\gamma)$ & $B^o\to\pi^+\pi^-$ \& $B_s\to K^+K^-$ & $\surd$ & $\surd$& & 
$\surd$ \\
$\sin(2\chi)$ & $B_s\to J/\psi\eta',$ $J/\psi\eta$ & & &$\surd$ &$\surd$\\
$\sin(2\beta)$ & $B^o\to J/\psi K_s$ & & & & \\
$\cos(2\beta)$ &  $B^o\to J/\psi K^o$, $K^o\to \pi\ell\nu$  & & & & \\
$\cos(2\beta)$ &  $B^o\to J/\psi K^{*o}$ \& $B_s\to J/\psi\phi$  & & & 
&$\surd$ \\
$x_s$  & $B_s\to D_s^+\pi^-$ & $\surd$ & & &$\surd$\\
$\Delta\Gamma$ for $B_s$ & $B_s\to  J/\psi\eta'$, $ D_s^+\pi^-$, $K^+K^-$ &
$\surd$ & $\surd$ & $\surd$ & $\surd$ \\
\hline
\end{tabular}
\end{center}
\end{table*}

I will focus my discussion on the experimental techniques most suitable for
a thorough investigation of beauty and charm decays. 
Table \ref{cp} summarizes the measurements that need to be done in beauty 
physics in
order to overconstrain the CKM matrix. The first measurements of some of
these 
quantities will be performed by the several new or upgraded experiments
pursuing $b$ physics in the next few years. In particular the \epm\ 
experiments, CLEO, BaBar and Belle should see CP violation in 2000. CLEO
is sensitive to direct CP violation in rare decays. BaBar and Belle should also
see CP violation in $B\dec \psi K_s$, where CDF has already 
seen a hint \cite{cdfbeta} . CDF and D0 are scheduled to turn on in 2001 and
are
both planning to pursue some of these measurements. HERA-B should also turn
on
in this time scale. Moreover ATLAS and CMS are planning to study CP violation
in $B\dec \psi K_s$ when LHC turns on.

None of these experiments  is capable of completing the physics program
summarized in Table~\ref{cp}. The \epm\ facilities are limited in 
statistical
accuracy for several measurements due to the low cross section for \bbar\ 
production. Furthermore, they cannot address $B_s$ physics. On the other
hand, the current, upgraded high \pt\ hadron collider experiments 
have a rudimentary particle identification system and do not have  
calorimeters suitable for the study of final states involving 
$\gamma$'s and $\pi ^o$'s with high resolution and efficiency. 
Furthermore they have limited triggers. Thus the need to design dedicated 
experiments to address beauty and
charm at hadronic machines has emerged as a crucial need to complete this 
physics program.

It is often customary to characterize heavy quark production in hadron
collisions with the two variables $p_t$ and $\eta$. The latter variable is 
defined as:
\begin{equation}
\eta = -ln\left(\tan\left({\theta/2}\right)\right),
\end{equation}
where $\theta$ is the angle of the particle with respect to the beam
direction.
According to QCD based calculations of $b$ quark production, the $b$'s are
produced ``uniformly" in $\eta$ and have a truncated transverse momentum, 
$p_t$, spectrum, characterized by a mean value approximately equal to the $B$
mass \cite{artuso}. 

\begin{figure}[htb]
\vspace{-.9cm}
\epsfig{figure=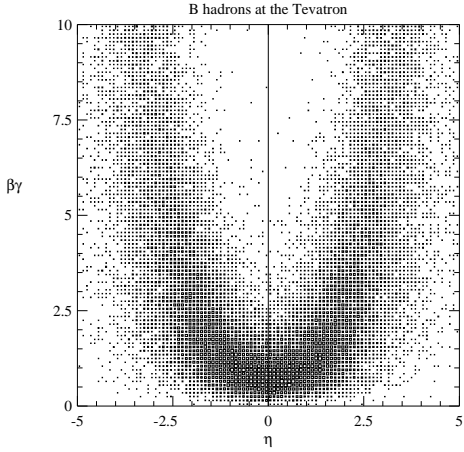,width=2.6in}
\caption{\label{n_vs_eta}  The  $\beta\gamma$ of the  $B$ 
 versus $\eta$  for the Tevatron.}
\end{figure}

There is a strong correlation between the $B$ momentum and $\eta$. Shown in
Fig.~\ref{n_vs_eta} is the $\beta\gamma$ of the $B$ hadron versus $\eta$.
It can clearly be seen that near $\eta=0$, $\beta\gamma\approx 1$, while
at larger values of $|\eta |$, $\beta\gamma$ can easily reach values of ~6.
This is important because the higher boost in the forward direction makes it 
easier to use vertices displaced from the primary interaction points as $b$
decay selection criteria.

\begin{figure}[htb]
\centerline{\epsfig{figure=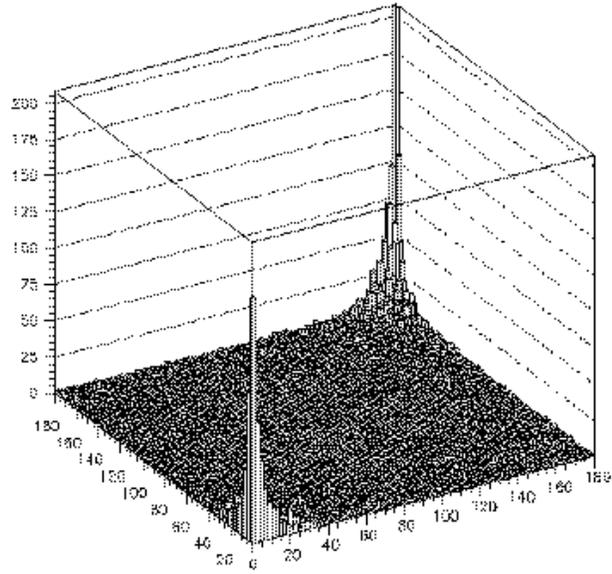,height=3.in}}
\caption{\label{bbar}The production angle (in degrees) for the hadron
containing a $b$ quark plotted versus the production angle for a hadron
containing $\bar{b}$ quark. (For the Tevatron.)}
\end{figure}

Fig.~\ref{bbar} shows another key feature of the $b\overline{b}$ production
at hadron colliders that favors the forward geometry: the production
angles of the hadron containing the $b$ quark is plotted versus the
production
angle of the hadron containing the $\bar{b}$ quark according to the Pythia
generator (for the Tevatron). There is a very strong
correlation in the forward (and backward) direction: when the $B$ is forward
the $\overline{B}$ is also forward. This correlation is not present in the
central region (near zero degrees). By instrumenting a relative small region
of angular phase space, a large number of $b\bar{b}$ pairs can be detected.
Furthermore the $B$'s populating the forward and backward regions have large
values of $\beta\gamma$. 

\begin{figure}[hbt]
\begin{center}
\epsfig{figure=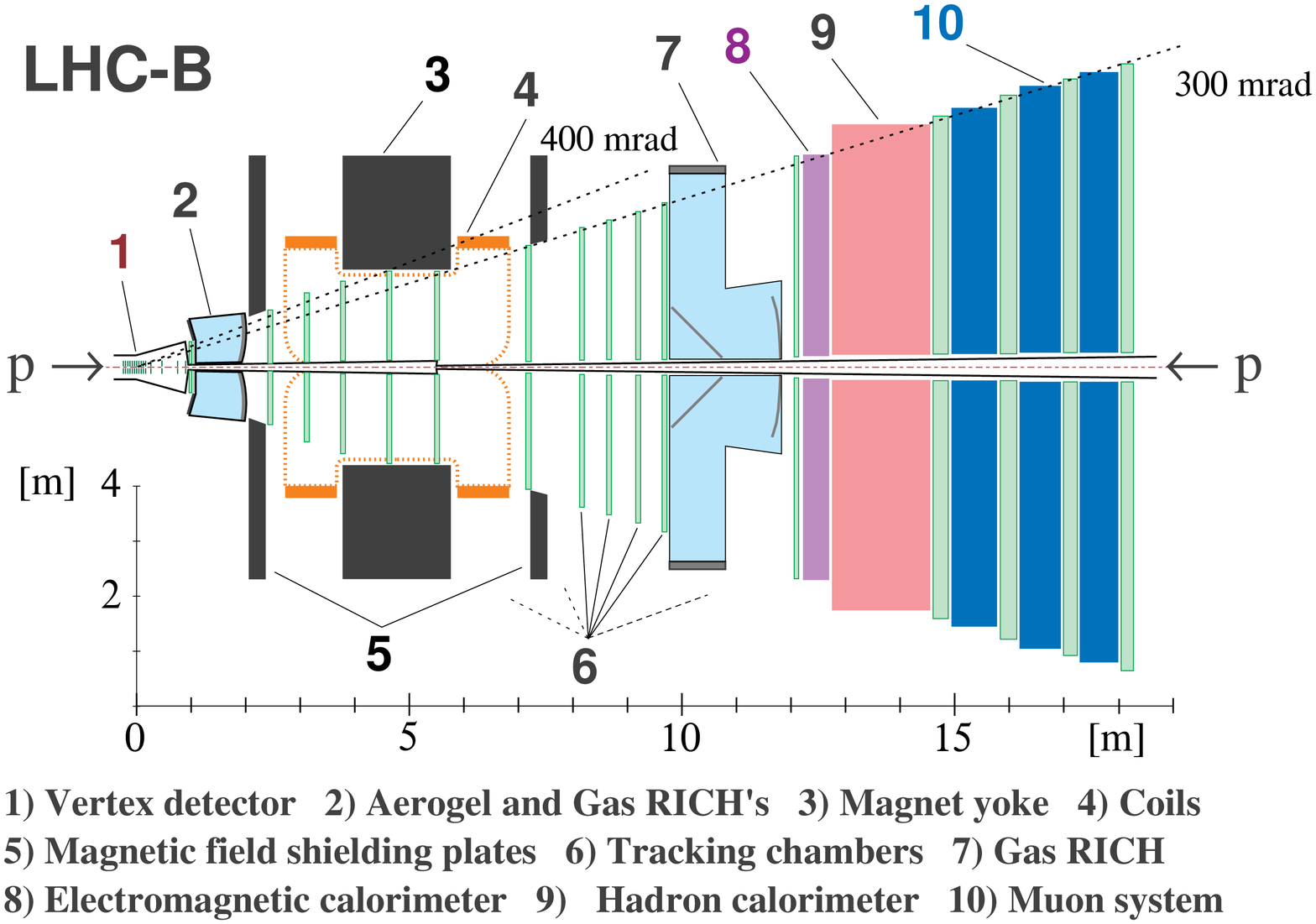,width=3in}
\epsfig{figure=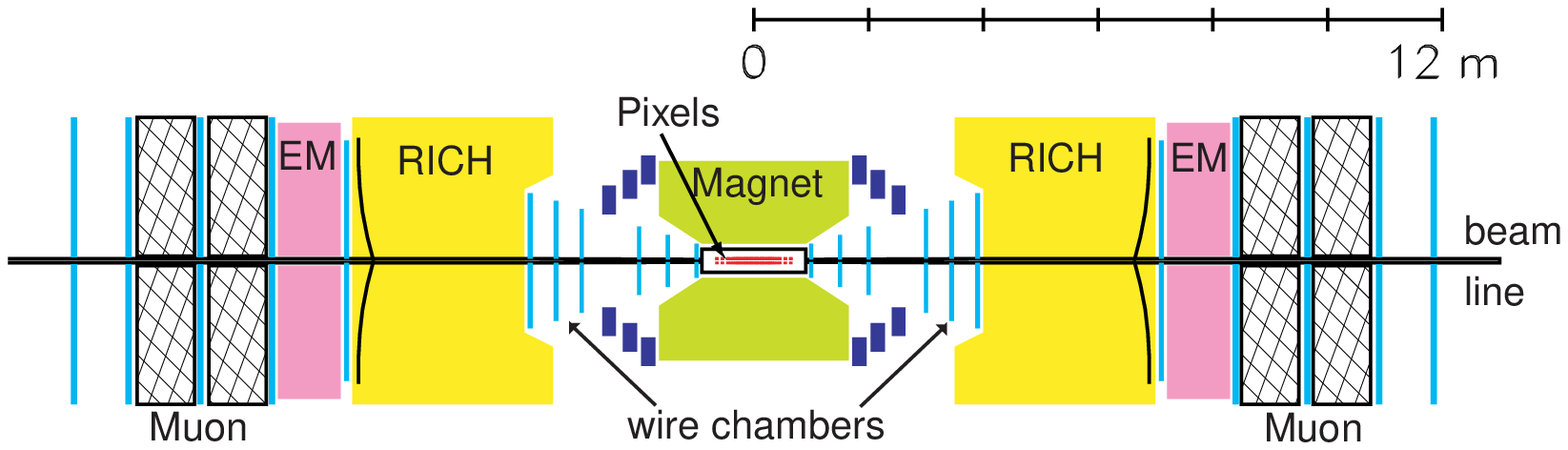,width=3in}
\caption{ Schematic view of the LHCb (a) and BTeV (b) detectors}
\label{btev-fig}
\end{center}
\end{figure}

The conclusions that we can draw from these data is that dedicated $B$ 
experiments at hadron colliders are
specially suited to pursue this physics with the highest prospects of
success.
 Two such experiments have been proposed, LHCb at the LHC and BTeV at
Fermilab. These detectors are shown in Fig.~\ref{btev-fig} a) and b) 
respectively. LHCb has been approved, while BTeV is an approved R\&D project; 
they have been invited to submit a proposal
in the summer of 2000. Both these experiments
exploit the high cross section for beauty production at hadronic machines and 
the advantages of the forward
geometry to achieve optimal performance. LHCb exploits the $\approx$ 5 
times higher cross
section at LHC compared to the Tevatron, due to the higher CM energy.

Both experiments feature excellent hadron identification. However, they
differ in their implementation of some key detector components. BTeV is 
planning to build a state of the art vertex detector based on high
granularity pixel sensors that are used in the first level trigger.  LHCb is 
considering microstrip silicon trackers whose information is used only at a 
higher trigger level. Furthermore BTeV has proposed an excellent crystal 
calorimeter based on PbWO$_4$, whereas LHCb will use a much more modest 
Shaslik calorimeter. The difference in the quality of the electromagnetic 
calorimeter is reflected by the difference in the projected sensitivities
 for the decay $B^o\dec \rho^{\pm}\pi^{\mp}$, 
where LHCb claims an efficiency of $1.7\times 10^{-4}$ \cite{lhcb-tdr}, 
whereas BTeV projects and efficiency of $(1.0\pm 0.2)$\% \cite{btev-ptdr}. 
Taking into account the difference in production cross section, the ratio 
between the projected yields per year in the 
two experiments favor BTeV with a ratio 12:1 with respect to 
LHCb \cite{btev-ptdr}.

\begin{table}[hbt]
\begin{center}
\caption{\label{phyreach} Sensitivies for BTeV running for one year and 
${\cal L}=2\times 10^{32}\ cm^{-2}s^{-1}$ and LHCb. The LHCb sensitivity
numbers have been taken from their Technical Design Report \cite{lhcb-tdr}
and the BTeV numbers from their Preliminary Technical Design 
Report \cite{btev-ptdr}. }
\begin{tabular}{lll}
\br
Measurement & BTeV  & LHCb \\
\hline
$x_s$ & $>80$ & $>74$ \\
CP asy ($B^o\dec\pi^+\pi^-$) & $\pm 0.032$ & $\pm 0.034$\\
$\gamma (D_sK)\dagger$ & $\pm 11^{\circ}$ & $\pm 6.1^{\circ } - 
\pm 13^{\circ }$\\
$\gamma (D^o K)$ & $\pm 13^{\circ}$ & ~~  \\
$\gamma (D^o K^{\star })$ & ~~ & $\pm 10^{\circ}$  \\
\br
\end{tabular}
\end{center}
$\dagger$ Both simulations assume $x_s=20$. The range of LHCb sensitivity 
corresponds to different 
assumptions on the strong and weak phases between
the amplitudes $<f|\overline{B}>$ and $<f|B>$. 
\end{table}

Table~\ref{phyreach} summarizes the expected physics reach of BTeV and LHCb
for
some key measurements in a year of running.  Their competitive effort will
insure a detailed and precise exploration of $b$ and $c$ decays and one of
the most thorough tests of the Standard Model in a sector where there is a 
stronger motivation for new physics.

\section{Conclusions}
The examples discussed in this paper show that much progress
has been made in the field of flavour physics. However, we have not 
 accomplished our main goals. The new
experiments being planned to do high precision studies of beauty and charm, 
BTeV and LHCb, have the opportunity of providing crucial information to
address
the puzzle of flavour and CP violation. In parallel, the study of rare $K$ 
decays can uncover 
complementary information on CP violation and  ATLAS and CMS
will explore the mechanism of electroweak symmetry breaking and maybe 
find some evidence for supersymmetry. In parallel, the experimental progress
will be undoubtely matched by a refinement of theoretical tools. 
Thus the mystery of flavour may be finally uncovered.
 
\section{Acknowledgements}
I would like to thank the organizers of this conference for
a very enjoyable and interesting conference. While preparing this manuscript, 
I have benefited from several interesting discussions with Sheldon
 Stone, Tomasz Skwarnicki, Zoltan Ligeti, Ikaros Bigi, Matthias Neubert
and Amarjit Soni.

\end{document}